\documentclass[letter]{aa} 

\usepackage{graphicx}
\usepackage{txfonts}
\usepackage[colorlinks = true,
            linkcolor = blue,
            urlcolor  = black,
            citecolor = cyan,
            anchorcolor = black]{hyperref}

\newcommand {\kms} {\,{\rm km\,s}^{-1}}
\newcommand{\hs}{\hspace*{5pt }}

\newcommand {\kpc} {\,{\rm kpc}}
\newcommand {\Mpc} {\,{\rm Mpc}}

\newcommand {\kmsMpc} {\,{\rm km\,s}^{-1}\,{\rm \Mpc}^{-1}}

\newcommand {\msun}{\,{\rm M}_\odot}
\newcommand {\mlsun}{\,{\rm M}_\odot/{\rm L}_\odot}

\newcommand{\Gyr}{\,{\rm Gyr}}

\newcommand{\bba}{$^{\scriptstyle 3\mathrm{D}}$B{\sc arolo}}
\def\apjl{ApJ}%
\def\apjs{ApJS}%
\def\ao{Appl.~Opt.}%
\def\aap{A\&A}%
\begin{document} 
\renewcommand{\figureautorefname}{Fig.\!}
\title{The angular momentum of disc galaxies at z=1}
\titlerunning{The angular momentum of disc galaxies at z=1}
\authorrunning{A. Marasco et al.}

   \author{A. Marasco\inst{1,2},
          F. Fraternali\inst{1},
          L. Posti\inst{1},
          M. Ijtsma\inst{1},
          E.\,M. Di Teodoro\inst{3},
          \and
          T. Oosterloo\inst{1,2}
          }

    \institute{Kapteyn Astronomical Institute, University of Groningen, P.O. Box 800, 9700 AV Groningen, the Netherlands\\
                  \email{marasco@astro.rug.nl}
         \and
         	 ASTRON, Netherlands Institute for Radio Astronomy, Oude Hoogeveensedijk 4, 7991 PD, Dwingeloo, The Netherlands
	\and
		Research School of Astronomy and Astrophysics - The Australian National University, Canberra, ACT, 2611, Australia
             }

   \date{Received ; accepted}

 
\abstract
{
We investigate the relation between stellar mass ($M_\star$) and specific stellar angular momentum ($j_\star$), or `Fall relation', for a sample of 17 isolated, regularly rotating disc galaxies at $z\!\sim\!1$.
All galaxies have a) rotation curves determined from H$\alpha$ emission-line data; b) HST imaging in optical and infrared filters; c) robust determinations of their stellar masses.
We use HST images in f814w and f160w filters, roughly corresponding to rest-frames B and I bands, to extract surface brightness profiles for our systems.
We robustly bracket $j_\star$ by assuming that rotation curves beyond the outermost H$\alpha$ rotation point stay either flat or follow a Keplerian fall-off.
By comparing our measurements with those determined for disc galaxies in the local Universe, we find no evolution in the Fall relation in the redshift range $0\!<\!z\!<\!1$, regardless of the band used and despite the uncertainties in the stellar rotation curves at large radii.
This result holds unless stellar masses at $z\!=\!1$ are systematically underestimated by $\gtrsim50\%$. Our findings are compatible with expectations based on a $\Lambda$CDM cosmological framework and support a scenario where both the stellar Tully-Fisher and mass-size relations for spirals do not evolve significantly in this redshift range.
}

\maketitle
%

\section{Introduction}
The stellar mass $M_\star$ and angular momentum $J_\star$ of a galaxy are two fundamental properties that are related qualitatively to the amount of material in the system, its rotational speed and its size.
Observationally, as firstly derived by \citet{Fall83}, $M_\star$ and specific angular momentum $j_\star\!=\!J_\star/M_\star$ are related as $j_\star\!\propto\!M_\star^\alpha$ ($\alpha\sim2/3$) with a normalisation that depends on the galaxy morphological type, increasing from early to late-type systems \citep{RF12,FR13}.

This `Fall' relation represents a key benchmark for models of galaxy evolution in a $\Lambda$CDM framework.
While tidal torque theory \citep{Peebles69,EfstathiouJones79} defines precise relations between the virial mass of halos and their specific angular momentum, which is also that of the baryonic matter, the mode by which the latter is redistributed within the stellar disc with time depends on the uncertain details of the galaxy evolution process \citep[e.g.][]{MMW,Dutton+12,Kravtsov13,Posti+18a}.
Early numerical studies suffered from dramatic angular momentum losses and strongly underpredicted the $j_\star$ of galaxies at all $M_\star$ \citep[e.g.][]{KatzGunn91,NavarroSteinmetz00}, but the situation has steadily improved in the years thanks to a better numerical resolution and to more effective stellar feedback implementations \citep{Governato+10,Genel+15,Lagos+17}.
It still holds, though, that realistic models of galaxy evolutions - analytical or numerical - must aim at reproducing the Fall relation in the local Universe \emph{and} its evolution with cosmic time.

The $j_\star-M_\star$ relation in the local Universe has been recently re-studied by \citet{Posti+18b} using a sample of 92 isolated, regularly rotating disc galaxies from the Spitzer Photometry and Accurate Rotation Curves \citep[SPARC,][]{SPARC} database, revealing a remarkably tight relation between the two quantities over four orders of magnitude in $M_\star$. 
The advent of modern IFU, like the KMOS and MUSE instruments on ESO/VLT, unveiled gas and stellar kinematics for galaxies at higher redshifts and opened the possibility to determine the evolution of the $j_\star\!-\!M_\star$ relation with time.
\citet{Harrison+17} studied a sample of 586 H$\alpha$-detected star-forming galaxies at $0.6\!<\!z\!<\!1$ from the KMOS Redshift One Spectroscopic Survey \citep[KROSS,][]{KROSS}, finding a power law $j_\star-M_\star$ relation with the same slope as in the local Universe, but with an offset of $0.2\!-\!0.3$ dex towards lower $j_\star$ which they explained as due to the smaller disc sizes.
Similar results were obtained by \citet{Swinbank+17} using KMOS and MUSE data for a sample of 405 galaxies at $z\!\sim\!0.84$, and by \citet{Burkert+16} for $360$ galaxies at $0.8\!<\!z\!<\!2.6$.
In contrast, \citet{Contini+16} studied a smaller sample of 28 galaxies at $0.2\!<\!z\!<\!1.4$ with MUSE, finding no evolution in the Fall relation in this redshift range for galaxies with large enough rotational support.

The reliability of $j_\star$ measurements depends on how accurately rotation curves and surface brightness profiles can be determined.
\citet[][hereafter TFM16]{DiTeodoro+16} selected 18 isolated, main-sequence disc galaxies at $z\!\sim\!1$ with intermediate inclination and high-quality H$\alpha$ data from the KROSS and the KMOS$^{\rm 3D}$ \citep{Wisnioski+15} surveys, and derived their H$\alpha$ rotation curve and velocity dispersion profiles using the tilted-ring code \bba\ \citep{Barolo}.
This approach, based on the modelling of the entire emission-line datacube, virtually bypasses any beam-smearing problem associated to low spatial resolution of the data and thus allows us to break the degeneracy between rotation velocity and velocity dispersion.
TFM16 concluded that the H$\alpha$ kinematical properties of these $z\!\sim\!1$ systems are analogous to those measured at $z\!=\!0$, with no evidence for additional dispersion support as previously claimed \citep[e.g.][]{Epinat+12,Kassin+12b}.

In this Letter, we determine the $j_\star-M_\star$ relation for the galaxy sample of TFM16 by combining their accurate H$\alpha$ kinematical measurements with surface brightness profiles that we extract from HST images.

\begin{figure*}[tbh]
\includegraphics[width=\textwidth]{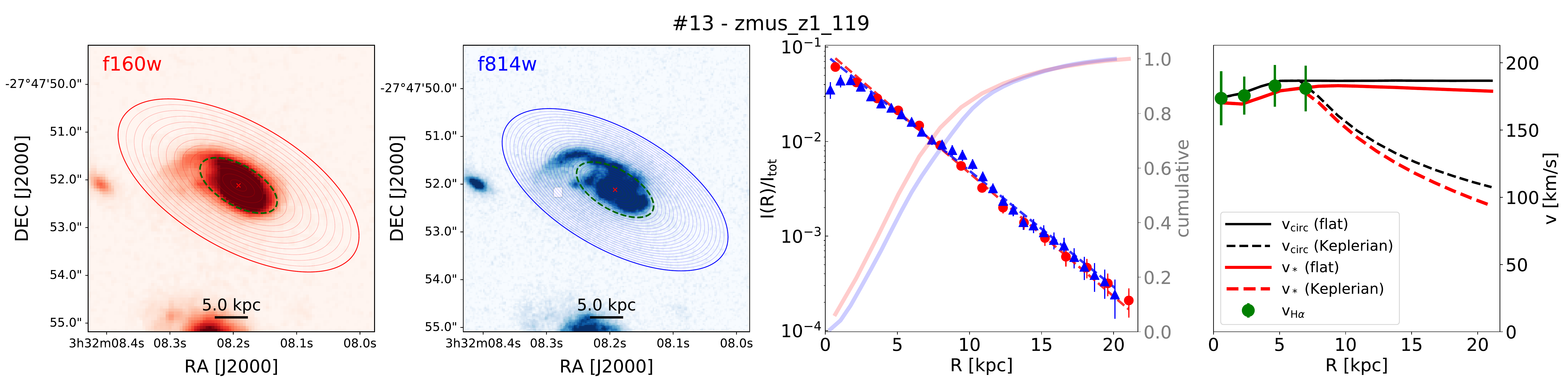}
\caption{Photometry and kinematics of galaxy zmus\_z1\_119. \emph{First panel}: HST image in f160w band (rest-frame I band). The concentric ellipses show the annuli where the surface brightness is computed. The green dashed ellipse shows the radius of the outermost H$\alpha$ rotation velocity point. The blank region in the f814w image has been masked by our sigma-rejection algorithm (see text). \emph{Second panel}: same as the first panel, but for the f814w band (rest-frame B band). \emph{Third panel}: surface brightness profiles in bands f160w (red circles) and f814w (blue triangles), normalised to the total light within the outermost ring. The thin dashed lines show exponential fits for $R\!>\!3.5\kpc$, the solid thick lines show the cumulative light profile. \emph{Fourth panel}: rotation curves. Points with error-bars show $\varv_{\rm H\alpha}$ as determined by TFM16. The black and the red solid (dashed) lines show our fiducial profiles for $\varv_{\rm circ}$ and $\varv_\star$ respectively, assuming a flat (Keplerian) extrapolation for the circular velocity.}
\label{profile_example}
\end{figure*} 

\section{Method}\label{method}
In an axisymmetric disc, the specific stellar angular momentum enclosed within a given radius $R$ is defined as
\begin{equation}\label{eq_js}
j_\star(<R) = \frac{\int_0^R\Sigma_\star(R^\prime)\,\varv_\star(R^\prime)\,R^{\prime 2}\,{\rm d}R^\prime}{\int_0^R\Sigma_\star(R^\prime)\,R^\prime\,{\rm d}R^\prime}
\end{equation}
where $\Sigma_\star(R^\prime)$ and $\varv_\star(R^\prime)$ are the surface density and the azimuthal velocity radial profiles of the stellar component.

In what follows, we focus on the task of computing $j_\star$ via eq.\,(\ref{eq_js}) for the galaxy sample of TFM16 and do not attempt our own determination of $M_\star$.
Instead, we use the stellar masses listed in Table 1 of TFM16, which come either from the Cosmic Assembly Near-infrared Deep Extragalactic Legacy Survey \citep[CANDELS,][]{Grogin+11,Koekemoer+11} as average values based on different techniques \citep{Santini+15}, or (for 5 galaxies) from the COSMOS \citep{cosmos} and 3D-HST \citep{Brammer+12} catalogues, which are derived by fitting stellar population synthesis templates to broad band photometry with the Fitting and Assessment of Synthetic Templates \citep[FAST,][]{Kriek+09} code.

The main properties of the galaxies studied in this work are reported in \autoref{tab:galaxies}.
Note that we follow the same galaxy name and IDs as in Table 1 of TFM16.
For consistency with TFM16, throughout this work we adopt a flat $\Lambda$CDM cosmology with $\Omega_{\rm m,0}\!=\!0.27$, $\Omega_{\Lambda,0}\!=\!0.73$  and $H_0\!=\!70\kmsMpc$.

\subsection{Surface brightness profiles} \label{sec_profiles}
We use optical/infrared surface brightness profiles $I(R)$ as a proxy for $\Sigma_\star(R)$ in eq.\,(\ref{eq_js}) under the assumption that light traces stellar mass.
We determine $I(R)$ by using publicly available HST images from CANDELS or, for systems 3, 8 and 17, from COSMOS.
No images were found for system 7 (zcos\_z1\_192), which we left out from our analysis.
We focus on two different filters, f160w and f814w, which at the redshift of $1$ roughly correspond to rest-frames I and B-bands, respectively.
According to \citet{Skelton+14}, images in f160w (f814w) filter have angular resolution of $\sim0.19\arcsec$ ($\sim0.10\arcsec$), corresponding to about $1.5\kpc$ ($0.8\kpc$) in physical units, $\sim3$ ($6$) times better than the KMOS IFU data.
In spite of the fact that the B-band is not the best tracer for stellar mass, we show below that our results are nearly independent of the band used, which strengthens our conclusions.

The procedure used to extract $I(R)$ from the images consists of several steps.
We first compute the central value $I_\mathrm{bkg}$ and width $\sigma_\mathrm{bkg}$ of the background noise distribution.
This is a crucial step, as an incorrect estimate for $I_{\rm bkg}$ affects the outer regions of the light profile, which may contain a significant fraction of $j_\star$.
We focus on the pixel intensity distribution in a region of the image outside the main galaxy and fit a Gaussian function to a window encompassing the mode of such distribution. 
The mean and the standard deviation of the best-fit Gaussian give $I_{\rm bkg}$ and $\sigma_{\rm bkg}$ respectively.
$I_{\rm bkg}$ is then subtracted from the image before any further analysis.

We then define an `optical' centre by using an iterative approach on the f160w image: we compute an initial intensity-weighted centroid by using all pixels within a circle centred at the galaxy coordinates given by TFM16 and with a radius of $\sim5\arcsec$, and progressively shrink the circle and re-center it to the newly computed centroid until convergence is reached.
The final centre coordinates are used for both bands and are reported in \autoref{tab:galaxies}.
We notice that a) optical and kinematic centres are in good agreement with each other (see Fig.\,2 in TFM16); b) our results are fairly robust against small ($\lesssim0.3\arcsec$) off-centring, given that most $j_\star$ is locked in the galaxy outskirts.

To extract $I(R)$, we consider a series of concentric annuli, centred on the optical centre, spaced out by one resolution element ($\sim\!3$ pixels) and with constant axis-ratio and orientation given by the inclination and position angle determined by TFM16.
As in \citet{Schombert+07}, a sigma-rejection algorithm is used to find and mask pixels contaminated by foreground stars or external galaxies. 
The surface brightness of each ring is computed as the mean intensity of the unmasked pixels, while the uncertainty associated to such measurement, $\delta I$, is given by max$(\sigma_{\rm bkg},\sigma)/\sqrt{N}$, being $\sigma$ the rms intensity of the unmasked pixels and $N$ the number of resolution elements in that ring.
We proceed ring by ring until the signal-to-noise $I / \delta I$ drops below a value of $2.5$.
Finally, we fit the profile with an exponential function in order to derive the scale length $R_{\rm d}$ of the disc.
In order to avoid contamination from bars/bulges, we exclude from the fit a region with $R\!<\!R_{\rm sph}$.
We obtain satisfying fits by choosing $R_{\rm sph}\!=\!3.5\kpc$ for all galaxies but we stress that, contrary to other works \citep[e.g.][]{Swinbank+17,Harrison+17}, in this analysis $R_{\rm d}$ does not enter directly in the computation of $j_\star$.
We find $2\!\lesssim\!R_{\rm d}\!\lesssim\!6\kpc$, as reported in \autoref{tab:galaxies}.

The first three panels in \autoref{profile_example} demonstrate our procedure for system zmus\_z1\_119.
Note that the optical profiles extend much further than the H$\alpha$ emission, as shown by the outermost solid and dashed ellipses in the leftmost panels of \autoref{profile_example}.
We also note the similarity between the normalised profiles in the two bands considered (third panel). 
The cumulative light profiles flattens out in both bands, which corroborates the validity of our procedure.
This is the case for all the galaxies in our sample, as we show in \autoref{fig:allprofiles}.

\subsection{Stellar rotation curves}\label{sec:vstar}
The computation of eq.\,(\ref{eq_js}) requires measurements for the \emph{stellar} rotation curves, $\varv_\star(R)$, that cover the full extent of the stellar discs.
At $z\!=\!1$, such measurements are very challenging \citep[for a recent attempt see][]{Bezanson+18}.
Fortunately, from the study of TFM16 we have access to H$\alpha$ rotation curves, $\varv_{\rm H\alpha}(R)$, although limited to the innermost (star-forming) regions.
Thus we must a) convert $\varv_{\rm H\alpha}$ to $\varv_\star$ via an asymmetric drift correction; b) make assumptions for the shape of the rotation curve beyond the outermost measured velocity point.

The computation of $\varv_\star$ is made in three steps: we first convert $\varv_{\rm H\alpha}$ to a circular velocity $\varv_{\rm circ}$ up to the outermost H$\alpha$ radius, then we extrapolate $\varv_{\rm circ}$ up to the outermost optical radius (see below), and finally we convert the whole $\varv_{\rm circ}$ profile to a $\varv_\star$ profile.
More quantitatively, we have
\begin{equation} \label{eq_vs}
\varv_\star^2 = \varv_{\rm circ}^2 - \varv_{\rm AD,\star}^2 =  \varv_{\rm H\alpha}^2 +  \varv_{\rm AD,H\alpha}^2 - \varv_{\rm AD,\star}^2
\end{equation}
being $\varv_{\rm AD,H\alpha}$ and $\varv_{\rm AD,\star}$ the asymmetric drift corrections for the gas and the stellar components. 
Following \citet{Meurer+96}, and assuming that galaxies have constant scale heights, these can be generically written as 
\begin{equation} \label{eq_AD}
\varv_{\rm AD}^2(R) = - R \left(\frac{\sigma_z(R)}{\beta}\right)^2 \frac{\partial \ln \left(\Sigma(R) \sigma_z(R)^2\right)} {\partial R}
\end{equation}
where $\Sigma(R)$ is the surface density profile of the component considered, and $\beta$ is defined as $\sigma_z/\sigma_R$, being $\sigma_z$ and $\sigma_R$ the vertical and radial components of the velocity dispersion\footnote{As it is often done in the literature, we have implicitly assumed that $\sigma_R\!=\!\sigma_\phi$ and that the off-diagonal elements of the velocity dispersion tensor are negligible.}.

To determine $\varv_{\rm AD,H\alpha}$ we assume isotropy ($\beta\!=\!1$) and a constant $\sigma_z$ set to the values listed by TFM16 in their Table 1. 
For further simplification, we assume $\Sigma_{{\rm H}\alpha}$ to follow an exponential profile with scale-length equal to that of the stellar disc determined in f814w band (see Section \ref{sec_profiles}).
For $\varv_{\rm AD,\star}$ we use a) the surface brightness profiles determined in the f160w and f814w bands as a proxy for $\Sigma$; b) an exponentially decreasing $\sigma_z$ profile with e-folding length given by $2 R_{\rm d}$ and central dispersion given by $\sigma_z(0)=(0.248\pm0.038)\times \varv_{2.2R_{\rm d}}$, being $\varv_{2.2R_{\rm d}}$ the circular velocity measured at $R\!=\!2.2R_{\rm d}$ \citep{Martinsson+13}.
In order to avoid unrealistically small values for $\sigma_z$ at large radii, we further impose a stellar velocity dispersion floor\footnote{The exact value of this floor has little impact on our results.} of $15\kms$.
Finally, we assume $0.5\!<\!\beta\!<\!1.0$ in order to account for the uncertainty in the velocity anisotropy in the error budget (see Section \ref{sec_error}).

For the extrapolation of $\varv_{\rm circ}$ we consider two scenarios: either the rotation velocity remains constant at the value set by the last measured point, or it follows a Keplerian fall-off.
The former represents a typical case for a late-type galaxy at $z\!=\!0$, while the latter is extreme and sets a conservative lower limit on $j_\star$.

As an example, in the rightmost panel of \autoref{profile_example} we show the `fiducial' $\varv_{\rm circ}$ and $\varv_\star$ profiles derived for system zmus\_z1\_119.
As we discuss in Section \ref{sec_error}, several realisations contribute to produce these fiducial profiles, each may differ markedly from those shown here.
Clearly, the type of extrapolation adopted dominates the uncertainty on the rotation curve (difference between solid and dashed lines), while the overall impact of the asymmetric drift corrections is small (difference between black and red lines).
For this reason, the two types of extrapolation will be treated separately in our study.

\subsection{Fiducial values and error budget}\label{sec_error}
Given the many sources of uncertainty, we adopt a Monte-Carlo approach to estimate `fiducial' values and associated errors on $j_\star$.
Our approach consists of producing $5\times10^4$ random realisations of the following quantities for any given galaxy: the H$\alpha$ rotation curve $\varv_{\rm H\alpha}(R)$, the H$\alpha$ velocity dispersion $\sigma_{\rm H\alpha}$, the stellar surface brightness profile $I(R)$, the central vertical dispersion for the stars $\sigma_z(0)$, and the stellar dispersion anisotropy $\beta$.
For the randomisation of $\varv_{\rm H\alpha}(R)$ and $\sigma_{\rm H\alpha}$ we adopt Gaussian uncertainties based on the error-bars determined by TFM16.
For $I(R)$ we use Gaussian uncertainties $\delta I$ computed as discussed in Section \ref{sec_profiles}.
For $\sigma_z(0)$ we use the formal (Gaussian) error determined by \citet{Martinsson+13} on their relation $\sigma_z(0)\!=\!(0.248\pm0.038)\times \varv_{2.2R_{\rm d}}$, where both $R_{\rm d}$ and $\varv_{2.2R_{\rm d}}$ are now random quantities depending on the realisation of the brightness and velocity profiles.
Finally, $\beta$ is randomly extracted from a uniform distribution between 0.5 and 1.

For each galaxy we compute eq.\,(\ref{eq_vs}) and (\ref{eq_js}) in all random realisations, and use the median and half the difference between the 84th and the 16th percentiles as our fiducial measurements and $1\sigma$ uncertainties associated to them, respectively.
The computation is done four times in total: once for each band separately and, for a given band, once for each extrapolation of the rotation curve (flat or Keplerian).

\section{Results}\label{results}
\begin{figure}[tbh]
\centering
\includegraphics[width=0.5\textwidth]{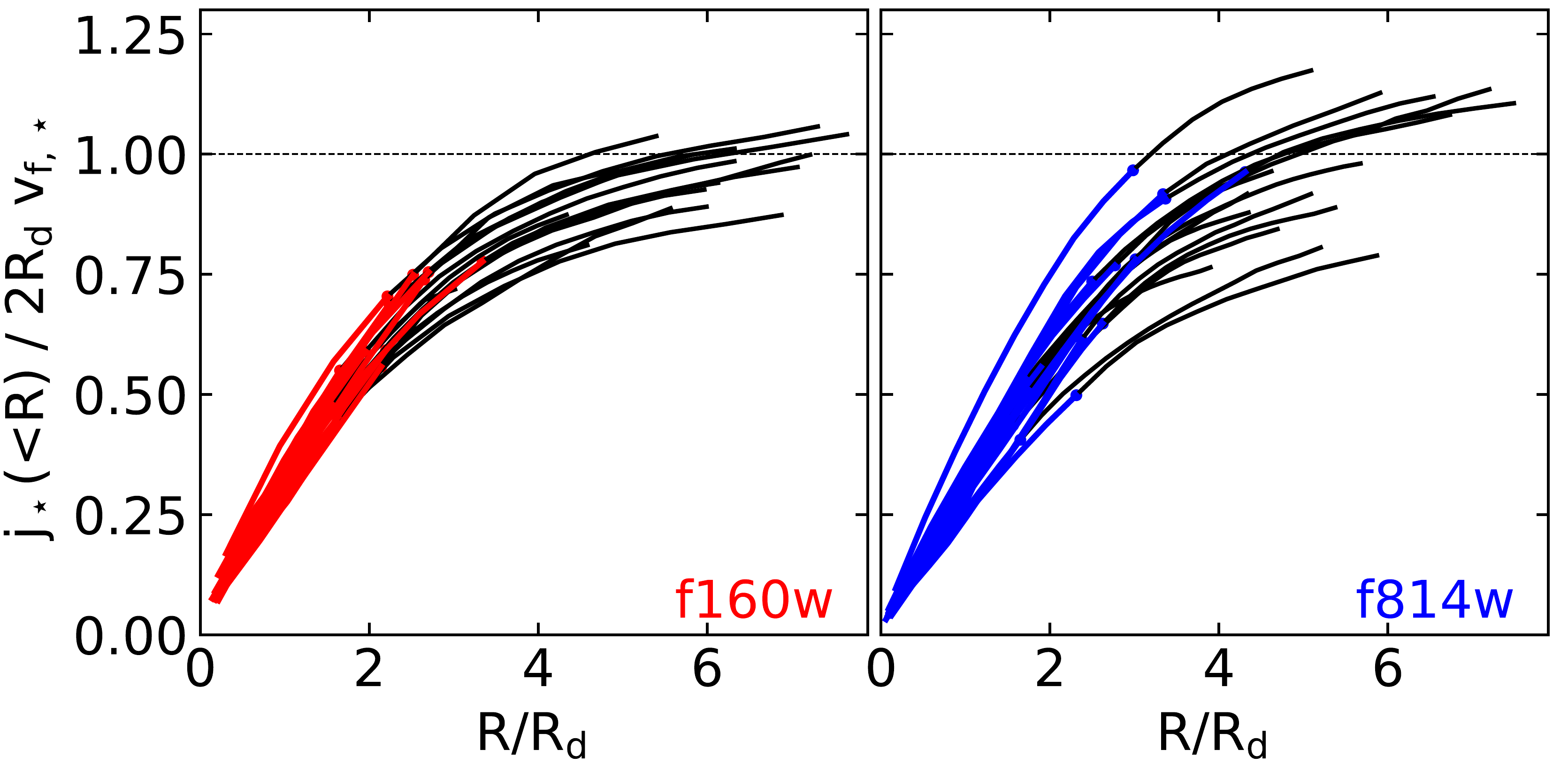}
\caption{Cumulative $j_\star$ profiles for our 17 galaxies in the f160w (left panel) and f814w (right panel) bands. Coloured solid lines show the innermost regions where rotation velocities from H$\alpha$ data are available, while solid black lines show the extrapolation for a flat rotation curve. All profiles are either converging or have fully converged.}
\label{js_cumulative}
\end{figure}

\begin{figure*}[tbh]
\centering
\includegraphics[width=0.85\textwidth]{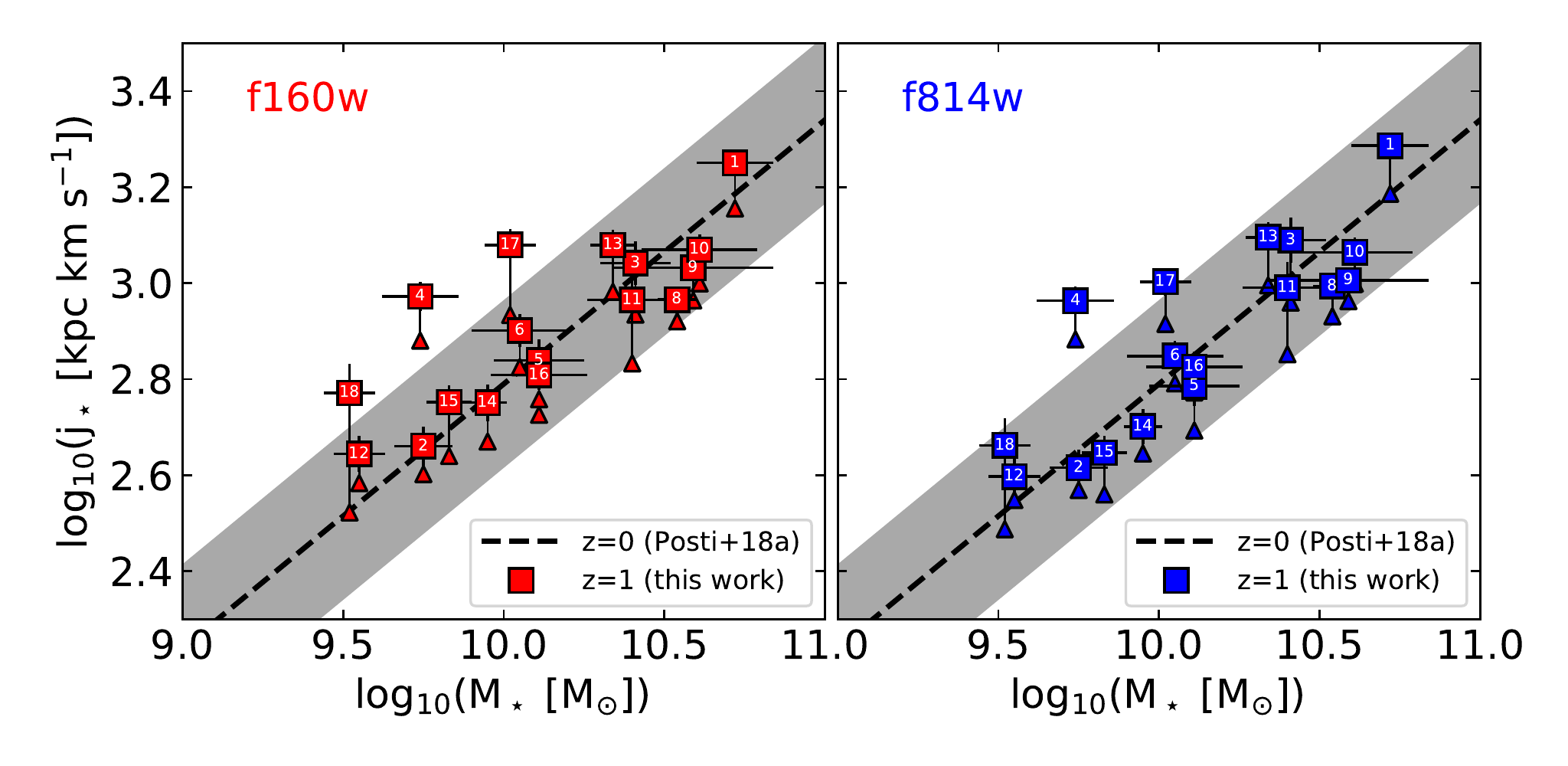}
\caption{Fall relation for our sample of disc galaxies at $z\!\sim\!1$. The \emph{left} (\emph{right}) panel shows the results derived using the f160w (f814w) HST band. The squared symbols show the $j_\star$ determined for a flat rotation curve extrapolation, while the upwards triangles represent lower limits derived from the assumption of a Keplerian fall-off. Numbers correspond to galaxy IDs from \autoref{tab:galaxies}. The black-dashed line and the grey-shaded region show the Fall relation and its intrinsic scatter as determined by \citet{Posti+18b} for disc galaxies in the local Universe from the SPARC data of \citet{SPARC}.}
\label{Fall_z1}
\end{figure*}
\autoref{js_cumulative} shows the cumulative $j_\star$ computed via  eq.\,(\ref{eq_js}) for all systems in our sample assuming a flat extrapolation for $\varv_{\rm circ}$.
All profiles shown are normalised by $2 R_{\rm d} \varv_{{\rm f},\star}$ (where $\varv_{{\rm f},\star}$ is maximum $\varv_\star$ in the extrapolated region of the curve), corresponding to the integrated $j_\star$ for an exponential disc with scale-length $R_{\rm d}$ and constant rotational speed $\varv_{{\rm f},\star}$.
The left (right) panel shows the results for the f160w (f814w) band. 
As expected, profiles in the f160w band show less scatter and a more uniform behaviour with respect to those in the f814w band, with a cleaner convergence towards the unity.
Following \citet{Posti+18b}, we have verified that all 17 systems satisfy in both bands the criteria for a `converging' $j_\star$ profile, $\Delta j_\star/j_\star\!<\!0.1$ and $\Delta{\rm log}j_\star/\Delta{\rm log}R<0.5$, where $\Delta$ is measured using the last two annuli of the profile.
Trivially, convergence is ensured also for the Keplerian extrapolation.
Note that none of the systems would pass the convergence test if we limited our study to their innermost regions traced by the H$\alpha$ emission, represented by the coloured portion of the curves shown in \autoref{js_cumulative}.

The Fall relation for our $z\!=\!1$ galaxies is presented in \autoref{Fall_z1} (points with error-bars) and compared to that determined in the local Universe by \citet{Posti+18b}\footnote{As shown by \citet{FR18}, this relation is in excellent agreement with those determined previously by \citet{FR13} and \citet{OG14} for disc-dominated galaxies.} (grey-shaded region) for the SPARC sample ($\log(j_\star)\!=\!0.55(\log(M_\star/\msun)-11)+3.34$, with a perpendicular scatter of 0.17 dex. 
Squared symbols (upwards triangles) represent the median $j_\star$ computed for a flat (Keplerian) extrapolation for the rotation curves.
Upwards triangles have been further shifted downwards by $1\sigma$ in order to represent strict lower limits.
Remarkably, all $z\!=\!1$ measured points sit comfortably on the $z\!=\!0$ relation of \citep{Posti+18b} with little dependence on the band adopted.
We reiterate that corrections for asymmetric drift (Section \ref{sec:vstar}) have a very small impact on our measurements, and stress that even our lower limits are typically enclosed within the scatter of the $z\!=\!0$ relation.

By assuming that the slope of the Fall relation does not vary in the redshift range considered, from our fiducial $j_\star$ measurements in f160w (f814w) band we infer a growth in the zero-point of the relation by $0.05$ ($0.03$) dex from $z\!=\!0$ to $z\!=\!1$.
This positive offset is mainly driven by systems 4, 17 and 18.
We test whether this measurement is compatible with zero offset by randomly extracting $10^5$ sets of 17 points in the $(j_\star,M_\star)$ plane using the relation of \citet{Posti+18b} as a probability distribution, taking into account both the intrinsic scatter and the typical error-bars of our $z\!=\!1$ data. We then compute the zero-point for each set, obtaining a distribution with a standard deviation of $\sim0.05$ dex, similar to the offset in the data.
This indicates that our measurements are compatible with \emph{no evolution} in the zero-point of the Fall relation for disc galaxies in the redshift range considered.

\section{Discussion}\label{discussion}
Our findings indicate that disc galaxies move along a well defined $j_\star-M_\star$ sequence while evolving from $z\!=\!1$ to $z\!=\!0$. 
As stellar mass grows, $j_\star$ increases by an amount defined by the Fall relation.

We stress that, while stellar masses for our $z\!=\!1$ sample come from broadband SED fitting, the SPARC sample (on which the $z\!=\!0$ Fall relation is based) adopts a constant M$_\star$/L$_{[3.6]}$ of $0.5\mlsun$ \citep{SPARC}.
It is unclear whether or not these two methods are compatible: while SED fitting techniques applied to nearby galaxies give masses corresponding to lower M$_\star$/L$_{[3.6]}$ \citep[$\sim\!0.1\!-\!0.3\mlsun$; see][]{Ponomareva+18,Hunt+18}, at a higher $z$ this may be no longer the case.
An in-depth analysis of these methods is beyond the purpose of this work, but a hint may come from the fact that our $z\!=\!1$ galaxies follow the same stellar Tully-Fisher (TF) relation as those in the local Universe (TFM16): were our stellar masses significantly underestimated, the zero-point of the TF would \emph{increase} with redshift, against all theoretical expectations \citep[e.g][]{Somerville+08,Ubler+17,Ferrero+17}.
Our measurements remain compatible with no evolution in the Fall relation unless stellar masses at $z\!=\!1$ ($z\!=\!0$) are systematically under(over)-estimated by $\gtrsim\!0.15\!-\!0.20$ dex ($\sim50\%$).

Our results appear to be in tension with those of \citet{Harrison+17}, who studied the H$\alpha$ kinematics of 586 star-forming galaxies at $z\!=\!0.6-1$ from the KROSS survey and concluded that, at a given $M_\star$, these galaxies have a deficiency of $\sim0.2-0.3$ dex in $j_\star$ with respect to their local counterpart.
To understand this discrepancy we cross-matched our sample with that of Harrison et al., finding 12 systems in common.
These galaxies do not occupy a preferred position in the ($M_\star,j_\star$) plane, thus are well representative for the overall population of rotationally dominated systems at $z\!=\!1$.
This excludes that our sample is biased towards high-$j_\star$ systems and indicates that the difference between the two works lies in the method.
We derive rotation velocity and velocity dispersion profiles via a 3D modelling of the H$\alpha$ datacubes, while Harrison et al. use major-axis $\varv_{\rm H\alpha}$ and $\sigma_{\rm H\alpha}$ profiles, an approach that often leads to underestimate the former and overestimate the latter \citep[e.g.][]{Barolo}. 
Also, they adopt an approximate estimator for $j_\star$ which relies on accurate measurements of disc sizes (see their eq.\,(5)).
By comparing the properties of the 12 galaxies in common we found that, while stellar masses are similar, our velocities and disc sizes are larger by about $0.08$ and $0.14$ dex respectively, leading to a $0.2$ dex larger $j_\star$, similar to the quoted deficiency with respect to the $z\!=\!0$ spirals.
Analogous arguments apply to the results of \citet{Swinbank+17}.

One may wonder whether our results must be expected on the basis of simple, first-principles models.
A straightforward, empirical approach is to consider $j_\star\propto R_{\rm d}\,\varv_{\rm f}$ and relate the evolution of $j_\star$ with that observed in the mass-size (MS) relation $M_\star\!-\!R_{\rm d}$ and in the stellar TF relation $M_\star\!-\!\varv_{\rm f}$.
Observationally, the evolution of both these relations is highly debated.
Our results support a scenario where both relations do not significantly change between $z\!=\!0$ and $z\!=\!1$, in agreement with the findings of \citet{Conselice+05}, \citet{Miller+11} and TFM16 for the TF relation and with those of \citet{Barden+05} for the MS relation of disc galaxies.
More recent measurements \citep{vanderWel+14,Huang+17,Somerville+18} have reported only a modest evolution ($\sim0.1$ dex) in the MS relation of late-type galaxies for the ($M_\star,z$) range of interest.

A more theoretical approach is to consider galaxy evolution models in a cosmological framework.
Following \citet{Obreschkow+15}, the $j_\star$ of a galaxy with stellar mass $M_\star$ that reside in a virialized spherical halo at a given redshift $z$ can be written as
\begin{equation}\label{eq_theory}
j_\star \propto \lambda \, f_{\rm j} \, f_{\rm m}^{-2/3} \, H(z)^{-1/3} \, \Delta_{\rm c}(z)^{-1/6} M_\star^{2/3}
\end{equation}
where $\lambda$ is the halo spin parameter ($\sim$constant in a $\Lambda$CDM universe), $f_{\rm m}\!\equiv\!M_\star/M_{\rm h}$ is the stellar-to-halo mass ratio, $f_{\rm j}\!\equiv\!j_\star/j_{\rm h}$ is the stellar-to-halo specific angular momentum ratio, $H(z)$ is the Hubble parameter and $\Delta_c(z)$ is the halo over-density relative to the mean density of the universe.
Both $f_{\rm m}$ and $f_{\rm j}$ are, in principle, functions of $M_\star$ and $z$.
Eq.\,(\ref{eq_theory}) leads immediately to
\begin{equation}\label{eq_ratio}
\frac{j_\star(M_\star,z)}{j_\star(M_\star,0)} =  \left[\frac{f_{\rm j}(M_\star,z)}{f_{\rm j}(M_\star,0)}\right] \left[\frac{f_{\rm m}(M_\star,z)}{f_{\rm m}(M_\star,0)}\right]^{-2/3} \left[\frac{H(z)}{H(0)}\right]^{-1/3} \left[\frac{\Delta_{\rm c}(z)}{\Delta_{\rm c}(0)}\right]^{-1/6}
\end{equation}
where we have explicitly written the dependence on $(M_\star,z)$ of the various quantities.

The $f_{\rm m}(M_\star,z)$ in eq.\,(\ref{eq_theory}) and (\ref{eq_ratio}) is fully determined by assuming a stellar-to-halo mass relation (SHMR).
Here we use the SHMR of \citet{Moster+13}, which is well defined at all redshifts and is thought to be well representative for the local Universe \citep[e.g.][]{Katz+17}.
If we ignore for a moment any evolution in $f_{\rm j}$, the right-hand term of eq.\,(\ref{eq_ratio}) computed at $z\!=\!0.9$ (the mean redshift of our sample) gives values ranging from $\sim\!+0.08$ dex at $\log(M_\star/\msun)\!=\!9.5$ to $\sim\!-0.08$ dex at $\log(M_\star/\msun)\!=\!10.7$. 
Adopting the SHMR of \citet{Behroozi+13} leads to a slightly larger range ($\pm0.1$ dex).
Alternatively, we can assume $f_{\rm j}\propto f_{\rm m}^s$ as in a `biased collapse' formation model where stars form inside-out accordingly to the cooling time of the available gas reservoir, so that the angular momentum retention efficiency $f_{\rm j}$ correlates with the star formation efficiency $f_{\rm m}$ \citep[see][]{vandenBosch98,Fall02,RF12,Posti+18a} at all redshift.
If we adopt $s\!=\!0.68$ as determined by \citet{Posti+18a}, then in eq.\,(\ref{eq_theory}) $f_{\rm j} \, f_{\rm m}^{-2/3}\!\simeq\!1$ and $j_\star(0.9)/j_\star(0)\!\simeq\!0.79$ ($-0.10$ dex), constant over the interested $M_\star$ range.
These predicted values, although uncertain, are within the scatter of the Fall relation, indicating that the lack of a strong variation in the relation between $z\!=\!0$ and $\sim1$ is consistent with simple galaxy evolution models.

Finally, conclusions similar to ours have been drawn by \citet{Lagos+17} from the analysis of the EAGLE cosmological simulations in a full $\Lambda$CDM framework \citep{Schaye+15}.
They found that the mean $j_\star$ computed within the half-$M_\star$ radius of star-forming galaxies with $10\!<\!\log(M_\star/\msun)\!<\!10.5$ decreases only marginally ($\sim\!0.05$ dex, see their Fig.\,7) in the redshift range considered, while differences become more marked at higher $z$. We plan to test this prediction in the next future by applying our technique to galaxies at $z\!\sim\!2$.

\section{Conclusions}\label{conclusions}
In this Letter we have studied the $j_\star-M_\star$ `Fall' relation for a sample of 17 regularly rotating disc galaxies at $z\sim1$.
All galaxies in our sample have resolved H$\alpha$ kinematics from KMOS IFU data, H$\alpha$ rotation curves from \citet{DiTeodoro+16}, HST images in optical and infrared bands from CANDELS, and robust determinations for their stellar masses.
We have determined their surface brightness profiles from HST images in f160w and f814w filters (rest-frame I and B bands, respectively), corrected the H$\alpha$ rotation curves for asymmetric drift, and computed $j_\star$ by assuming that either rotation velocities stay constant beyond the outermost H$\alpha$ data point or follow a Keplerian fall-off.
The latter scenario provides a lower limit on $j_\star$.
All systems show converging $j_\star$ profiles, which makes them suitable for our study.

We have found that the Fall relation at this redshift is compatible with that determined by \citet{Posti+18b} for spirals in the local Universe, unless $z\!=\!1$ stellar masses derived via SED fitting have been underestimated by more than $\sim\!50\%$.
This implies that the Fall relation for disc galaxies has not evolved significantly in the last $\sim8\Gyr$.
Our findings are in line with expectations based on simple galaxy evolution models and with cosmological simulation in a $\Lambda$CDM framework, and support a scenario where both the stellar Tully-Fisher and the mass-size relations of disc galaxies do not evolve significantly between redshift $1$ and $0$.

\begin{acknowledgements}
The authors thank the referee for an insightful report, and are grateful to M. Fall for his comments and insights.
A.M. thanks F. Lelli for useful discussions.
L.P. acknowledges financial support from a VICI grant from the Netherlands Organisation for Scientific Research (NWO).
E.D.T. acknowledges the support of the Australian Research Council (ARC) through grant DP160100723.
\end{acknowledgements}


\bibliographystyle{aa} 
\bibliography{aa} 

\begin{thebibliography}{57}
\expandafter\ifx\csname natexlab\endcsname\relax\def\natexlab#1{#1}\fi

\bibitem[{{Barden} {et~al.}(2005){Barden}, {Rix}, {Somerville}, {Bell},
  {H{\"a}u{\ss}ler}, {Peng}, {Borch}, {Beckwith}, {Caldwell}, {Heymans},
  {Jahnke}, {Jogee}, {McIntosh}, {Meisenheimer}, {S{\'a}nchez}, {Wisotzki}, \&
  {Wolf}}]{Barden+05}
{Barden}, M., {Rix}, H.-W., {Somerville}, R.~S., {et~al.} 2005, \apj, 635, 959

\bibitem[{{Behroozi} {et~al.}(2013){Behroozi}, {Wechsler}, \&
  {Conroy}}]{Behroozi+13}
{Behroozi}, P.~S., {Wechsler}, R.~H., \& {Conroy}, C. 2013, \apj, 770, 57

\bibitem[{{Bezanson} {et~al.}(2018){Bezanson}, {van der Wel}, {Pacifici},
  {Noeske}, {Bari{\v s}i{\'c}}, {Bell}, {Brammer}, {Calhau}, {Chauke}, {van
  Dokkum}, {Franx}, {Gallazzi}, {van Houdt}, {Labb{\'e}}, {Maseda},
  {Mu{\~n}os-Mateos}, {Muzzin}, {van de Sande}, {Sobral}, {Straatman}, \&
  {Wu}}]{Bezanson+18}
{Bezanson}, R., {van der Wel}, A., {Pacifici}, C., {et~al.} 2018, \apj, 858, 60

\bibitem[{{Brammer} {et~al.}(2012){Brammer}, {van Dokkum}, {Franx},
  {Fumagalli}, {Patel}, {Rix}, {Skelton}, {Kriek}, {Nelson}, {Schmidt},
  {Bezanson}, {da Cunha}, {Erb}, {Fan}, {F{\"o}rster Schreiber}, {Illingworth},
  {Labb{\'e}}, {Leja}, {Lundgren}, {Magee}, {Marchesini}, {McCarthy},
  {Momcheva}, {Muzzin}, {Quadri}, {Steidel}, {Tal}, {Wake}, {Whitaker}, \&
  {Williams}}]{Brammer+12}
{Brammer}, G.~B., {van Dokkum}, P.~G., {Franx}, M., {et~al.} 2012, \apjs, 200,
  13

\bibitem[{{Burkert} {et~al.}(2016){Burkert}, {F{\"o}rster Schreiber}, {Genzel},
  {Lang}, {Tacconi}, {Wisnioski}, {Wuyts}, {Bandara}, {Beifiori}, {Bender},
  {Brammer}, {Chan}, {Davies}, {Dekel}, {Fabricius}, {Fossati}, {Kulkarni},
  {Lutz}, {Mendel}, {Momcheva}, {Nelson}, {Naab}, {Renzini}, {Saglia},
  {Sharples}, {Sternberg}, {Wilman}, \& {Wuyts}}]{Burkert+16}
{Burkert}, A., {F{\"o}rster Schreiber}, N.~M., {Genzel}, R., {et~al.} 2016,
  \apj, 826, 214

\bibitem[{{Conselice} {et~al.}(2005){Conselice}, {Bundy}, {Ellis}, {Brichmann},
  {Vogt}, \& {Phillips}}]{Conselice+05}
{Conselice}, C.~J., {Bundy}, K., {Ellis}, R.~S., {et~al.} 2005, \apj, 628, 160

\bibitem[{{Contini} {et~al.}(2016){Contini}, {Epinat}, {Bouch{\'e}},
  {Brinchmann}, {Boogaard}, {Ventou}, {Bacon}, {Richard}, {Weilbacher},
  {Wisotzki}, {Krajnovi{\'c}}, {Vielfaure}, {Emsellem}, {Finley}, {Inami},
  {Schaye}, {Swinbank}, {Gu{\'e}rou}, {Martinsson}, {Michel-Dansac},
  {Schroetter}, {Shirazi}, \& {Soucail}}]{Contini+16}
{Contini}, T., {Epinat}, B., {Bouch{\'e}}, N., {et~al.} 2016, \aap, 591, A49

\bibitem[{{Di Teodoro} \& {Fraternali}(2015)}]{Barolo}
{Di Teodoro}, E.~M. \& {Fraternali}, F. 2015, \mnras, 451, 3021

\bibitem[{{Di Teodoro} {et~al.}(2016){Di Teodoro}, {Fraternali}, \&
  {Miller}}]{DiTeodoro+16}
{Di Teodoro}, E.~M., {Fraternali}, F., \& {Miller}, S.~H. 2016, \aap, 594, A77

\bibitem[{{Dutton} \& {van den Bosch}(2012)}]{Dutton+12}
{Dutton}, A.~A. \& {van den Bosch}, F.~C. 2012, \mnras, 421, 608

\bibitem[{{Efstathiou} \& {Jones}(1979)}]{EfstathiouJones79}
{Efstathiou}, G. \& {Jones}, B.~J.~T. 1979, \mnras, 186, 133

\bibitem[{{Epinat} {et~al.}(2012){Epinat}, {Tasca}, {Amram}, {Contini}, {Le
  F{\`e}vre}, {Queyrel}, {Vergani}, {Garilli}, {Kissler-Patig}, {Moultaka},
  {Paioro}, {Tresse}, {Bournaud}, {L{\'o}pez-Sanjuan}, \& {Perret}}]{Epinat+12}
{Epinat}, B., {Tasca}, L., {Amram}, P., {et~al.} 2012, \aap, 539, A92

\bibitem[{{Fall}(1983)}]{Fall83}
{Fall}, S.~M. 1983, in IAU Symposium, Vol. 100, Internal Kinematics and
  Dynamics of Galaxies, ed. E.~{Athanassoula}, 391--398

\bibitem[{{Fall}(2002)}]{Fall02}
{Fall}, S.~M. 2002, in Astronomical Society of the Pacific Conference Series,
  Vol. 275, Disks of Galaxies: Kinematics, Dynamics and Peturbations, ed.
  E.~{Athanassoula}, A.~{Bosma}, \& R.~{Mujica}, 389--396

\bibitem[{{Fall} \& {Romanowsky}(2013)}]{FR13}
{Fall}, S.~M. \& {Romanowsky}, A.~J. 2013, \apjl, 769, L26

\bibitem[{{Fall} \& {Romanowsky}(2018)}]{FR18}
{Fall}, S.~M. \& {Romanowsky}, A.~J. 2018, ArXiv e-prints
  [\eprint[arXiv]{1808.02525}]

\bibitem[{{Ferrero} {et~al.}(2017){Ferrero}, {Navarro}, {Abadi}, {Sales},
  {Bower}, {Crain}, {Frenk}, {Schaller}, {Schaye}, \& {Theuns}}]{Ferrero+17}
{Ferrero}, I., {Navarro}, J.~F., {Abadi}, M.~G., {et~al.} 2017, \mnras, 464,
  4736

\bibitem[{{Genel} {et~al.}(2015){Genel}, {Fall}, {Hernquist}, {Vogelsberger},
  {Snyder}, {Rodriguez-Gomez}, {Sijacki}, \& {Springel}}]{Genel+15}
{Genel}, S., {Fall}, S.~M., {Hernquist}, L., {et~al.} 2015, \apjl, 804, L40

\bibitem[{{Governato} {et~al.}(2010){Governato}, {Brook}, {Mayer}, {Brooks},
  {Rhee}, {Wadsley}, {Jonsson}, {Willman}, {Stinson}, {Quinn}, \&
  {Madau}}]{Governato+10}
{Governato}, F., {Brook}, C., {Mayer}, L., {et~al.} 2010, \nat, 463, 203

\bibitem[{{Grogin} {et~al.}(2011){Grogin}, {Kocevski}, {Faber}, {Ferguson}, \&
  {Koekemoer}}]{Grogin+11}
{Grogin}, N.~A., {Kocevski}, D.~D., {Faber}, S.~M., {Ferguson}, H.~C., \&
  {Koekemoer}, A.~M. e.~a. 2011, \apjs, 197, 35

\bibitem[{{Harrison} {et~al.}(2017){Harrison}, {Johnson}, {Swinbank}, {Stott},
  {Bower}, {Smail}, {Tiley}, {Bunker}, {Cirasuolo}, {Sobral}, {Sharples},
  {Best}, {Bureau}, {Jarvis}, \& {Magdis}}]{Harrison+17}
{Harrison}, C.~M., {Johnson}, H.~L., {Swinbank}, A.~M., {et~al.} 2017, \mnras,
  467, 1965

\bibitem[{{Huang} {et~al.}(2017){Huang}, {Fall}, {Ferguson}, {van der Wel},
  {Grogin}, {Koekemoer}, {Lee}, {P{\'e}rez-Gonz{\'a}lez}, \&
  {Wuyts}}]{Huang+17}
{Huang}, K.-H., {Fall}, S.~M., {Ferguson}, H.~C., {et~al.} 2017, \apj, 838, 6

\bibitem[{{Hunt} {et~al.}(2018){Hunt}, {De Looze}, {Boquien}, {Nikutta},
  {Rossi}, {Bianchi}, {Dale}, {Granato}, {Kennicutt}, {Silva}, {Ciesla},
  {Relano}, {Viaene}, {Brandl}, {Calzetti}, {Croxall}, {Draine}, {Galametz},
  {Gordon}, {Groves}, {Helou}, {Herrera-Camus}, {Hinz}, {Koda}, {Salim},
  {Sandstrom}, {Smith}, {Wilson}, \& {Zibetti}}]{Hunt+18}
{Hunt}, L.~K., {De Looze}, I., {Boquien}, M., {et~al.} 2018, ArXiv e-prints
  [\eprint[arXiv]{1809.04088}]

\bibitem[{{Kassin} {et~al.}(2012){Kassin}, {Weiner}, {Faber}, {Gardner},
  {Willmer}, {Coil}, {Cooper}, {Devriendt}, {Dutton}, {Guhathakurta}, {Koo},
  {Metevier}, {Noeske}, \& {Primack}}]{Kassin+12b}
{Kassin}, S.~A., {Weiner}, B.~J., {Faber}, S.~M., {et~al.} 2012, \apj, 758, 106

\bibitem[{{Katz} {et~al.}(2017){Katz}, {Lelli}, {McGaugh}, {Di Cintio},
  {Brook}, \& {Schombert}}]{Katz+17}
{Katz}, H., {Lelli}, F., {McGaugh}, S.~S., {et~al.} 2017, \mnras, 466, 1648

\bibitem[{{Katz} \& {Gunn}(1991)}]{KatzGunn91}
{Katz}, N. \& {Gunn}, J.~E. 1991, \apj, 377, 365

\bibitem[{{Koekemoer} {et~al.}(2011){Koekemoer}, {Faber}, {Ferguson}, {Grogin},
  \& {Kocevski}}]{Koekemoer+11}
{Koekemoer}, A.~M., {Faber}, S.~M., {Ferguson}, H.~C., {Grogin}, N.~A., \&
  {Kocevski}, D.~D. e.~a. 2011, \apjs, 197, 36

\bibitem[{{Kravtsov}(2013)}]{Kravtsov13}
{Kravtsov}, A.~V. 2013, \apjl, 764, L31

\bibitem[{{Kriek} {et~al.}(2009){Kriek}, {van Dokkum}, {Labb{\'e}}, {Franx},
  {Illingworth}, {Marchesini}, \& {Quadri}}]{Kriek+09}
{Kriek}, M., {van Dokkum}, P.~G., {Labb{\'e}}, I., {et~al.} 2009, \apj, 700,
  221

\bibitem[{{Lagos} {et~al.}(2017){Lagos}, {Theuns}, {Stevens}, {Cortese},
  {Padilla}, {Davis}, {Contreras}, \& {Croton}}]{Lagos+17}
{Lagos}, C.~d.~P., {Theuns}, T., {Stevens}, A.~R.~H., {et~al.} 2017, \mnras,
  464, 3850

\bibitem[{{Lelli} {et~al.}(2016){Lelli}, {McGaugh}, \& {Schombert}}]{SPARC}
{Lelli}, F., {McGaugh}, S.~S., \& {Schombert}, J.~M. 2016, \aj, 152, 157

\bibitem[{{Martinsson} {et~al.}(2013){Martinsson}, {Verheijen}, {Westfall},
  {Bershady}, {Schechtman-Rook}, {Andersen}, \& {Swaters}}]{Martinsson+13}
{Martinsson}, T.~P.~K., {Verheijen}, M.~A.~W., {Westfall}, K.~B., {et~al.}
  2013, \aap, 557, A130

\bibitem[{{Meurer} {et~al.}(1996){Meurer}, {Carignan}, {Beaulieu}, \&
  {Freeman}}]{Meurer+96}
{Meurer}, G.~R., {Carignan}, C., {Beaulieu}, S.~F., \& {Freeman}, K.~C. 1996,
  \aj, 111, 1551

\bibitem[{{Miller} {et~al.}(2011){Miller}, {Bundy}, {Sullivan}, {Ellis}, \&
  {Treu}}]{Miller+11}
{Miller}, S.~H., {Bundy}, K., {Sullivan}, M., {Ellis}, R.~S., \& {Treu}, T.
  2011, \apj, 741, 115

\bibitem[{{Mo} {et~al.}(1998){Mo}, {Mao}, \& {White}}]{MMW}
{Mo}, H.~J., {Mao}, S., \& {White}, S.~D.~M. 1998, \mnras, 295, 319

\bibitem[{{Moster} {et~al.}(2013){Moster}, {Naab}, \& {White}}]{Moster+13}
{Moster}, B.~P., {Naab}, T., \& {White}, S.~D.~M. 2013, \mnras, 428, 3121

\bibitem[{{Navarro} \& {Steinmetz}(2000)}]{NavarroSteinmetz00}
{Navarro}, J.~F. \& {Steinmetz}, M. 2000, \apj, 538, 477

\bibitem[{{Obreschkow} \& {Glazebrook}(2014)}]{OG14}
{Obreschkow}, D. \& {Glazebrook}, K. 2014, \apj, 784, 26

\bibitem[{{Obreschkow} {et~al.}(2015){Obreschkow}, {Glazebrook}, {Bassett},
  {Fisher}, {Abraham}, {Wisnioski}, {Green}, {McGregor}, {Damjanov}, {Popping},
  \& {J{\o}rgensen}}]{Obreschkow+15}
{Obreschkow}, D., {Glazebrook}, K., {Bassett}, R., {et~al.} 2015, \apj, 815, 97

\bibitem[{{Peebles}(1969)}]{Peebles69}
{Peebles}, P.~J.~E. 1969, \apj, 155, 393

\bibitem[{{Ponomareva} {et~al.}(2018){Ponomareva}, {Verheijen}, {Papastergis},
  {Bosma}, \& {Peletier}}]{Ponomareva+18}
{Ponomareva}, A.~A., {Verheijen}, M.~A.~W., {Papastergis}, E., {Bosma}, A., \&
  {Peletier}, R.~F. 2018, \mnras, 474, 4366

\bibitem[{{Posti} {et~al.}(2018{\natexlab{a}}){Posti}, {Fraternali}, {Di
  Teodoro}, \& {Pezzulli}}]{Posti+18b}
{Posti}, L., {Fraternali}, F., {Di Teodoro}, E.~M., \& {Pezzulli}, G.
  2018{\natexlab{a}}, \aap, 612, L6

\bibitem[{{Posti} {et~al.}(2018{\natexlab{b}}){Posti}, {Pezzulli},
  {Fraternali}, \& {Di Teodoro}}]{Posti+18a}
{Posti}, L., {Pezzulli}, G., {Fraternali}, F., \& {Di Teodoro}, E.~M.
  2018{\natexlab{b}}, \mnras, 475, 232

\bibitem[{{Romanowsky} \& {Fall}(2012)}]{RF12}
{Romanowsky}, A.~J. \& {Fall}, S.~M. 2012, \apjs, 203, 17

\bibitem[{{Santini} {et~al.}(2015){Santini}, {Ferguson}, {Fontana}, {Mobasher},
  {Barro}, {Castellano}, {Finkelstein}, {Grazian}, {Hsu}, {Lee}, {Lee},
  {Pforr}, {Salvato}, {Wiklind}, {Wuyts}, {Almaini}, {Cooper}, {Galametz},
  {Weiner}, {Amorin}, {Boutsia}, {Conselice}, {Dahlen}, {Dickinson},
  {Giavalisco}, {Grogin}, {Guo}, {Hathi}, {Kocevski}, {Koekemoer},
  {Kurczynski}, {Merlin}, {Mortlock}, {Newman}, {Paris}, {Pentericci},
  {Simons}, \& {Willner}}]{Santini+15}
{Santini}, P., {Ferguson}, H.~C., {Fontana}, A., {et~al.} 2015, \apj, 801, 97

\bibitem[{{Schaye} {et~al.}(2015){Schaye}, {Crain}, {Bower}, {Furlong},
  {Schaller}, {Theuns}, {Dalla Vecchia}, {Frenk}, {McCarthy}, {Helly},
  {Jenkins}, {Rosas-Guevara}, {White}, {Baes}, {Booth}, {Camps}, {Navarro},
  {Qu}, {Rahmati}, {Sawala}, {Thomas}, \& {Trayford}}]{Schaye+15}
{Schaye}, J., {Crain}, R.~A., {Bower}, R.~G., {et~al.} 2015, \mnras, 446, 521

\bibitem[{{Schombert}(2007)}]{Schombert+07}
{Schombert}, J. 2007, ArXiv Astrophysics e-prints [\eprint{astro-ph/0703646}]

\bibitem[{{Scoville} {et~al.}(2007){Scoville}, {Aussel}, {Brusa}, {Capak},
  {Carollo}, {Elvis}, {Giavalisco}, {Guzzo}, {Hasinger}, {Impey}, {Kneib},
  {LeFevre}, {Lilly}, {Mobasher}, {Renzini}, {Rich}, {Sanders}, {Schinnerer},
  {Schminovich}, {Shopbell}, {Taniguchi}, \& {Tyson}}]{cosmos}
{Scoville}, N., {Aussel}, H., {Brusa}, M., {et~al.} 2007, \apjs, 172, 1

\bibitem[{{Skelton} {et~al.}(2014){Skelton}, {Whitaker}, {Momcheva}, {Brammer},
  \& {van Dokkum}}]{Skelton+14}
{Skelton}, R.~E., {Whitaker}, K.~E., {Momcheva}, I.~G., {Brammer}, G.~B., \&
  {van Dokkum}, P.~G. e.~a. 2014, \apjs, 214, 24

\bibitem[{{Somerville} {et~al.}(2008){Somerville}, {Barden}, {Rix}, {Bell},
  {Beckwith}, {Borch}, {Caldwell}, {H{\"a}u{\ss}ler}, {Heymans}, {Jahnke},
  {Jogee}, {McIntosh}, {Meisenheimer}, {Peng}, {S{\'a}nchez}, {Wisotzki}, \&
  {Wolf}}]{Somerville+08}
{Somerville}, R.~S., {Barden}, M., {Rix}, H.-W., {et~al.} 2008, \apj, 672, 776

\bibitem[{{Somerville} {et~al.}(2018){Somerville}, {Behroozi}, {Pandya},
  {Dekel}, {Faber}, {Fontana}, {Koekemoer}, {Koo}, {P{\'e}rez-Gonz{\'a}lez},
  {Primack}, {Santini}, {Taylor}, \& {van der Wel}}]{Somerville+18}
{Somerville}, R.~S., {Behroozi}, P., {Pandya}, V., {et~al.} 2018, \mnras, 473,
  2714

\bibitem[{{Stott} {et~al.}(2016){Stott}, {Swinbank}, {Johnson}, {Tiley},
  {Magdis}, {Bower}, {Bunker}, {Bureau}, {Harrison}, {Jarvis}, {Sharples},
  {Smail}, {Sobral}, {Best}, \& {Cirasuolo}}]{KROSS}
{Stott}, J.~P., {Swinbank}, A.~M., {Johnson}, H.~L., {et~al.} 2016, \mnras,
  457, 1888

\bibitem[{{Swinbank} {et~al.}(2017){Swinbank}, {Harrison}, {Trayford},
  {Schaller}, {Smail}, {Schaye}, {Theuns}, {Smit}, {Alexander}, {Bacon},
  {Bower}, {Contini}, {Crain}, {de Breuck}, {Decarli}, {Epinat}, {Fumagalli},
  {Furlong}, {Galametz}, {Johnson}, {Lagos}, {Richard}, {Vernet}, {Sharples},
  {Sobral}, \& {Stott}}]{Swinbank+17}
{Swinbank}, A.~M., {Harrison}, C.~M., {Trayford}, J., {et~al.} 2017, \mnras,
  467, 3140

\bibitem[{{{\"U}bler} {et~al.}(2017){{\"U}bler}, {F{\"o}rster Schreiber},
  {Genzel}, {Wisnioski}, {Wuyts}, {Lang}, {Naab}, {Burkert}, {van Dokkum},
  {Tacconi}, {Wilman}, {Fossati}, {Mendel}, {Beifiori}, {Belli}, {Bender},
  {Brammer}, {Chan}, {Davies}, {Fabricius}, {Galametz}, {Lutz}, {Momcheva},
  {Nelson}, {Saglia}, {Seitz}, \& {Tadaki}}]{Ubler+17}
{{\"U}bler}, H., {F{\"o}rster Schreiber}, N.~M., {Genzel}, R., {et~al.} 2017,
  \apj, 842, 121

\bibitem[{{van den Bosch}(1998)}]{vandenBosch98}
{van den Bosch}, F.~C. 1998, \apj, 507, 601

\bibitem[{{van der Wel} {et~al.}(2014){van der Wel}, {Franx}, {van Dokkum},
  {Skelton}, {Momcheva}, {Whitaker}, {Brammer}, {Bell}, {Rix}, {Wuyts},
  {Ferguson}, {Holden}, {Barro}, {Koekemoer}, {Chang}, {McGrath},
  {H{\"a}ussler}, {Dekel}, {Behroozi}, {Fumagalli}, {Leja}, {Lundgren},
  {Maseda}, {Nelson}, {Wake}, {Patel}, {Labb{\'e}}, {Faber}, {Grogin}, \&
  {Kocevski}}]{vanderWel+14}
{van der Wel}, A., {Franx}, M., {van Dokkum}, P.~G., {et~al.} 2014, \apj, 788,
  28

\bibitem[{{Wisnioski} {et~al.}(2015){Wisnioski}, {F{\"o}rster Schreiber},
  {Wuyts}, {Wuyts}, {Bandara}, {Wilman}, {Genzel}, {Bender}, {Davies},
  {Fossati}, {Lang}, {Mendel}, {Beifiori}, {Brammer}, {Chan}, {Fabricius},
  {Fudamoto}, {Kulkarni}, {Kurk}, {Lutz}, {Nelson}, {Momcheva}, {Rosario},
  {Saglia}, {Seitz}, {Tacconi}, \& {van Dokkum}}]{Wisnioski+15}
{Wisnioski}, E., {F{\"o}rster Schreiber}, N.~M., {Wuyts}, S., {et~al.} 2015,
  \apj, 799, 209

\end{thebibliography}

\appendix
\section{Supplementary material}
\autoref{tab:galaxies} lists the main properties of the 17 galaxies at $z\!\sim\!1$ analyzed in this work.
Galaxy names, IDs, redshifts and stellar masses are the same as those reported Table 1 of TFM16, while the other properties have been determined in this work.

As in \autoref{profile_example} for the case of zmus\_z1\_119, in \autoref{fig:allprofiles} we present the photometric and kinematic analysis for the other 16 galaxies of our sample.

\begin{sidewaystable*}
\caption{Properties of the 17 galaxies at $z\!\sim\!1$ analyzed in this work.}
\label{tab:galaxies} 
\centering
\def\arraystretch{1.3}
\begin{tabular}{rlcccccccccc}
\hline\hline\noalign{\vspace{5pt}}
\#& Name & R.A.\ (J2000)	& Dec.\ (J2000)	 & $z$ & $\log\frac{M_\star}{\msun}$ & $R_{\rm d}^{\rm\textcolor{red}{f160w}}$  & $R_{\rm d}^{\rm \textcolor{blue}{f814w}}$  & $j_{\star,\rm flat}^{\rm\textcolor{red}{f160w}}$ & $j_{\star,\rm flat}^{\rm\textcolor{blue}{f814w}}$ & $j_{\star,\rm Kep}^{\rm\textcolor{red}{f160w}}$ & $j_{\star,\rm Kep}^{\rm\textcolor{blue}{f814w}}$\\
	&		&$^\mathrm{h\hs m\hs s}$&$^{\circ}$\hs '\hs ''&&&\tiny{kpc}&\tiny{kpc}&\tiny{$\times10^2\kpc\kms$}&\tiny{$\times10^2\kpc\kms$}&\tiny{$\times10^2\kpc\kms$}&\tiny{$\times10^2\kpc\kms$}\\
\hspace{4pt} & (1) & (2) & (3) & (4) & (5) & (6) & (7) & (8) & (9) & (10) & (11)\\
\noalign{\smallskip}
\hline\noalign{\vspace{5pt}}

1  & gs3\_22005	&   	03 32 29.86  & -27 45 20.7 & 0.954   & $10.72\pm0.12$ 	& $5.1\pm0.4$ & $6.1\pm0.6$ & $17.80\pm0.92$ & $19.34\pm1.03$ & $15.09\pm0.80$ & $16.24\pm0.91$ \\
2  & hiz\_z1\_195 	&   	10 00 34.64  & +02 14 29.6 & 0.856  & $9.75\pm0.09$ 	& $1.9\pm0.2$ & $1.8\pm0.2$ & $4.58\pm0.43$ & $4.13\pm0.36$ & $4.40\pm0.41$ & $4.04\pm0.35$ \\
3  & hiz\_z1\_258	&   	10 01 05.65  & +01 52 57.7 & 0.838 & $10.41\pm0.11$ 	& $4.4\pm0.3$ & $3.8\pm0.2$ & $10.99\pm1.16$ & $12.28\pm1.32$ & $9.52\pm0.98$ & $10.13\pm1.08$ \\
4  & u3\_5138		&   	02 16 59.89  & -05 15 07.6 & 0.809   & $9.74\pm0.12$ 	& $4.1\pm0.2$ & $4.2\pm0.2$ & $9.37\pm0.66$ & $9.18\pm0.62$ & $8.15\pm0.58$ & $8.17\pm0.55$ \\
5  & u3\_14150	 	&   	02 16 58.00  & -05 12 42.6 & 0.896   & $10.11\pm0.14$ 	& $2.2\pm0.1$ & $1.9\pm0.1$ & $6.91\pm0.69$ & $6.09\pm0.58$ & $5.88\pm0.59$ & $5.41\pm0.51$ \\
6  & u3\_25160		&   	02 17 04.69  & -05 09 46.5 & 0.897   & $10.05\pm0.15$ 	& $3.0\pm0.2$ & $2.9\pm0.2$ & $7.98\pm0.63$ & $7.05\pm0.51$ & $7.25\pm0.58$ & $6.64\pm0.48$ \\
8  & zcos\_z1\_202	&   	10 00 53.39  & +01 52 40.9 & 0.841 & $10.54\pm0.06$ 	& $2.2\pm0.1$ & $2.1\pm0.1$ & $9.24\pm0.42$ & $9.87\pm0.47$ & $8.71\pm0.39$ & $8.94\pm0.41$ \\
9  & zcos\_z1\_690	&   	10 00 36.55  & +02 13 09.5 & 0.927 & $10.59\pm0.25$ 	& $2.4\pm0.1$ & $2.1\pm0.1$ & $10.76\pm0.67$ & $10.13\pm0.61$ & $9.79\pm0.60$ & $9.74\pm0.58$ \\
10& zcos\_z1\_692	&   	10 00 36.42  & +02 11 19.2 & 0.930 & $10.61\pm0.18$ 	& $3.1\pm0.2$ & $3.2\pm0.2$ & $11.73\pm0.85$ & $11.60\pm0.82$ & $10.72\pm0.76$ & $10.70\pm0.74$ \\
11& zmus\_z1\_21	&   	03 32 48.48  & -27 54 16.0 &  0.839 & $10.40\pm0.14$ 	& $3.3\pm0.2$ & $3.8\pm0.2$ & $9.23\pm1.12$ & $9.78\pm1.21$ & $7.70\pm0.94$ & $8.05\pm1.00$ \\
12& zmus\_z1\_86	&   	03 32 25.20  & -27 51 00.1  & 0.841 & $9.55\pm0.08$ 	& $1.9\pm0.1$ & $1.6\pm0.1$ & $4.42\pm0.37$ & $3.96\pm0.34$ & $4.19\pm0.37$ & $3.86\pm0.33$ \\
13& zmus\_z1\_119 	&	03 32 08.20  & -27 47 52.1  & 0.840 & $10.34\pm0.07$ 	& $3.3\pm0.2$ & $3.5\pm0.2$ & $12.01\pm0.86$ & $12.46\pm0.91$ & $10.30\pm0.73$ & $10.63\pm0.77$ \\
14& zmus\_z1\_125	&   	03 32 21.76  & -27 47 24.7  & 0.998 & $9.95\pm0.06$ 	& $1.7\pm0.1$ & $1.4\pm0.1$ & $5.64\pm0.50$ & $5.03\pm0.42$ & $5.12\pm0.46$ & $4.81\pm0.40$ \\
15& zmus\_z1\_129  &   	03 32 26.29  & -27 47 17.5  & 0.995 & $9.83\pm0.07$ 	& $2.3\pm0.1$ & $2.1\pm0.1$ & $5.65\pm0.46$ & $4.43\pm0.36$ & $4.77\pm0.42$ & $3.94\pm0.32$ \\
16& zmus\_z1\_166  &  	03 32 16.49  & -27 44 49.0  & 0.975 & $10.11\pm0.15$ 	& $2.0\pm0.1$  & $1.9\pm0.1$ &$6.44\pm0.47$ & $6.70\pm0.49$ & $6.17\pm0.45$ & $6.38\pm0.47$ \\
17& zmus\_z1\_217  &	03 32 20.53  & -27 40 58.8  & 0.895 & $10.02\pm0.08$ 	& $4.0\pm0.2$  & $3.9\pm0.2$ &$12.00\pm0.93$ & $10.08\pm0.67$ & $9.32\pm0.76$ & $8.80\pm0.58$ \\
18& zmvvd\_z1\_87  &	03 32 05.66  & -27 47 49.1  & 0.896 & $9.52\pm0.08$ 	& $3.3\pm0.2$  & $2.8\pm0.2$ &$5.91\pm0.84$ & $4.59\pm0.59$ & $4.04\pm0.81$ & $3.58\pm0.55$ \\

\noalign{\vspace{2pt}}\hline
\noalign{\vspace{5pt}}

\multicolumn{12}{p{0.92\textwidth}}{\textbf{Notes.} (1) Name adopted in the main survey (KROSS or KMOS$^{\rm 3D}$); (2)-(3) Celestial coordinates in J2000 as determined in this work; (4) Spectroscopic redshift; (5) Stellar masses from CANDELS or COSMOS/3D-HST (see text); (6)-(7) Disc scale-length in f160w/f814w band as determined in this work; (8)-(11) Stellar specific angular momentum in f160w/f814w band for a flat/Keplerian extrapolation of the circular velocity.}\\
\vspace*{15pt}
\end{tabular}
\end{sidewaystable*}

\begin{figure*}[tb]
\centering
\includegraphics[width=1.0\textwidth]{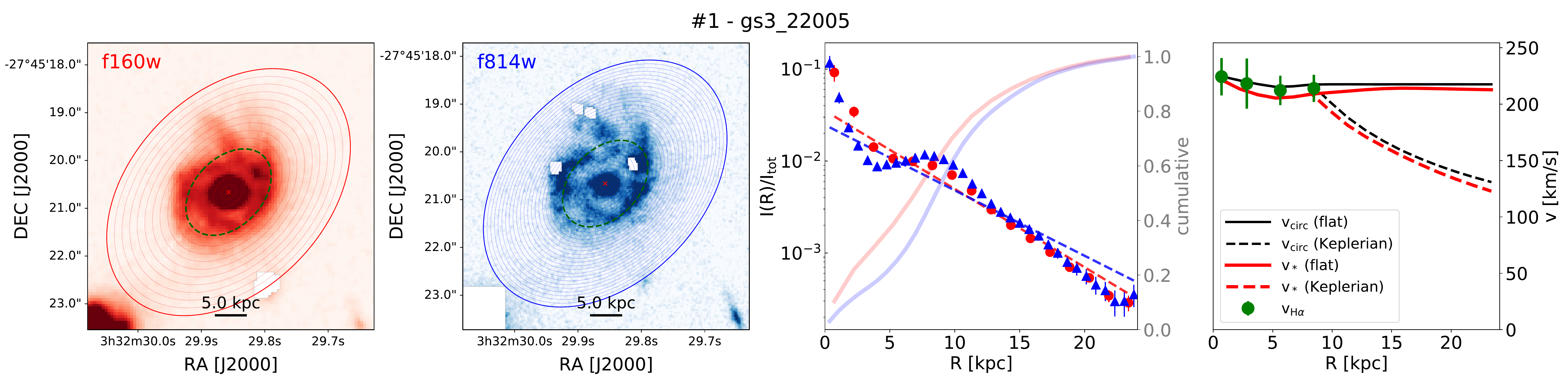}
\includegraphics[width=1.0\textwidth]{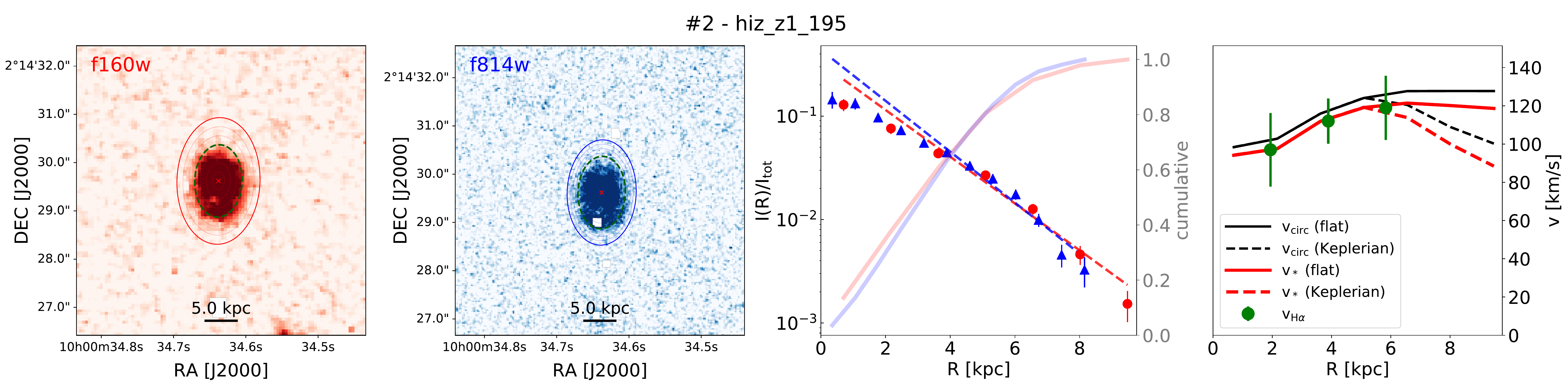}
\includegraphics[width=1.0\textwidth]{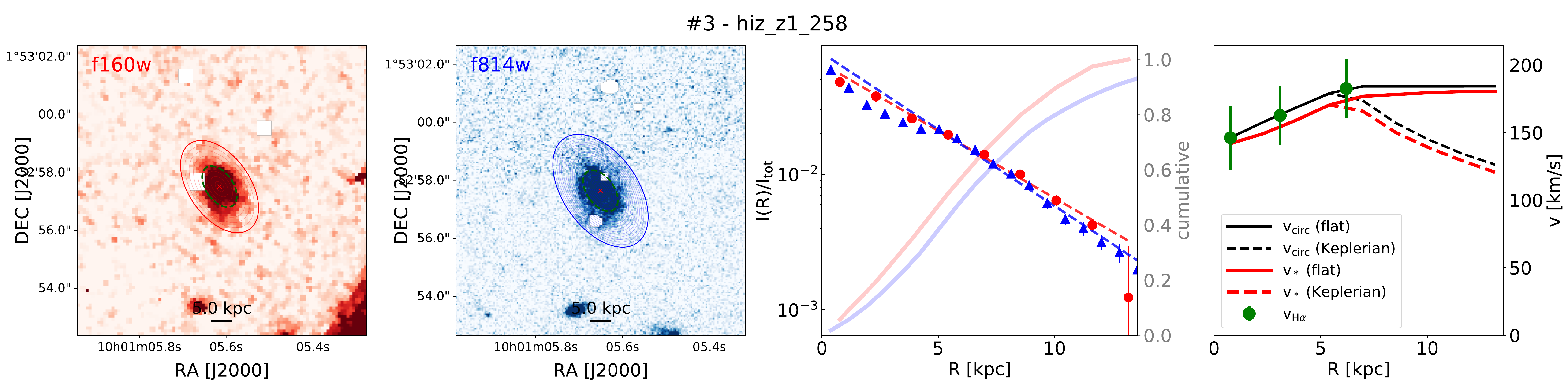}
\includegraphics[width=1.0\textwidth]{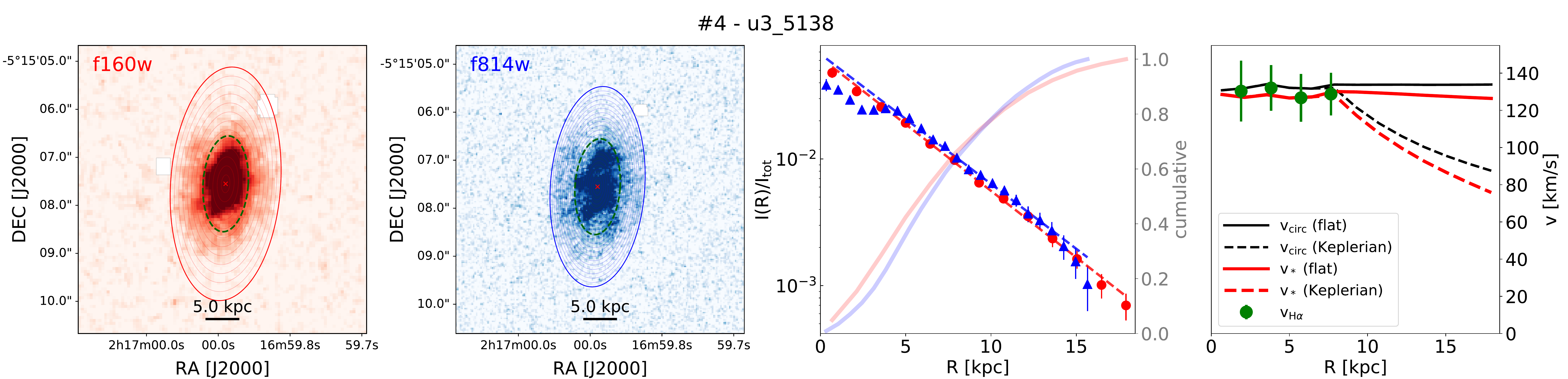}
\includegraphics[width=1.0\textwidth]{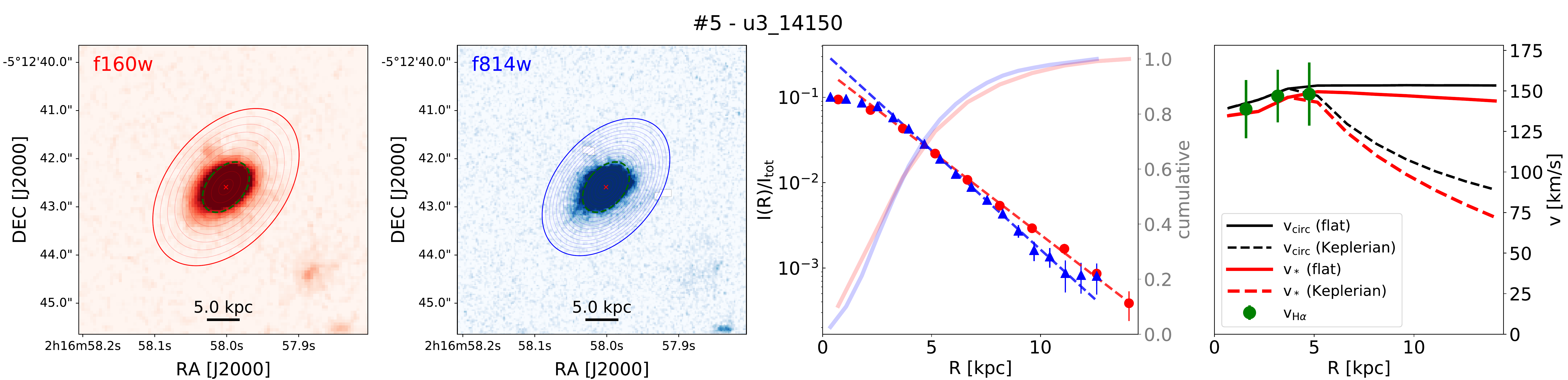}
\caption{Same as \autoref{profile_example}, but for the other 16 galaxies in our sample.}
\label{fig:allprofiles}
\end{figure*}
\addtocounter{figure}{-1}

\begin{figure*}[tb]
\centering

\includegraphics[width=1.0\textwidth]{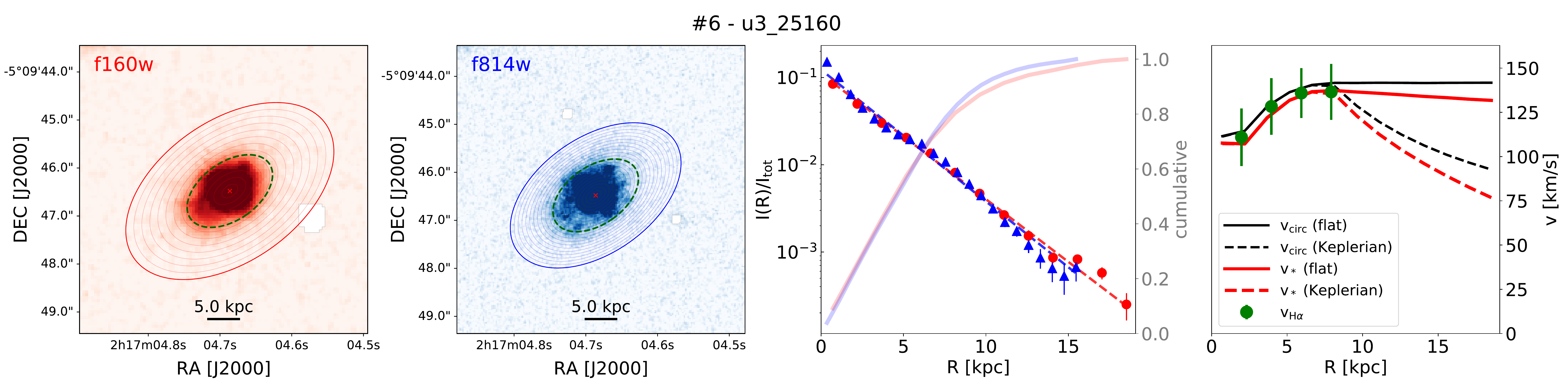}
\includegraphics[width=1.0\textwidth]{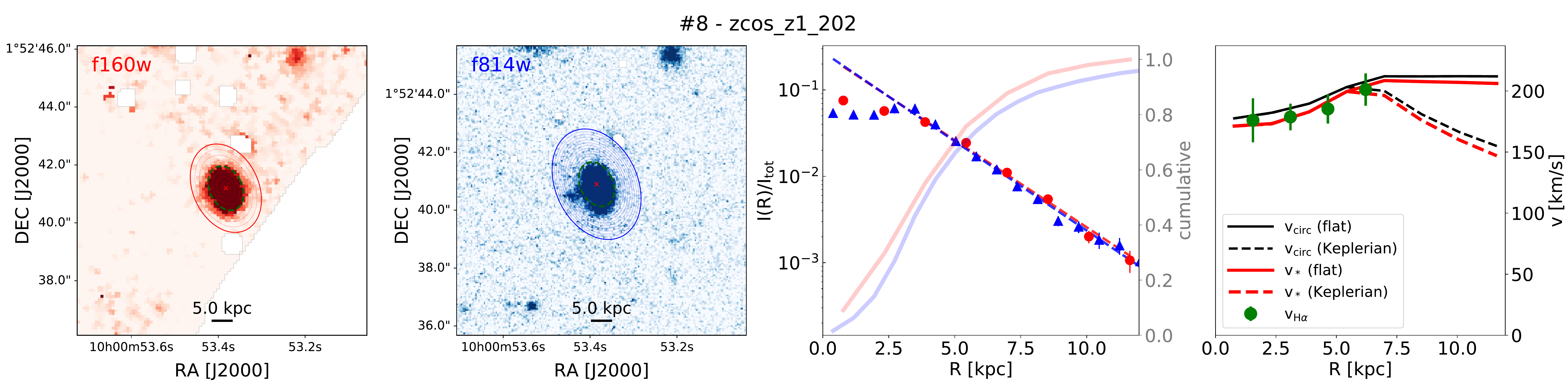}
\includegraphics[width=1.0\textwidth]{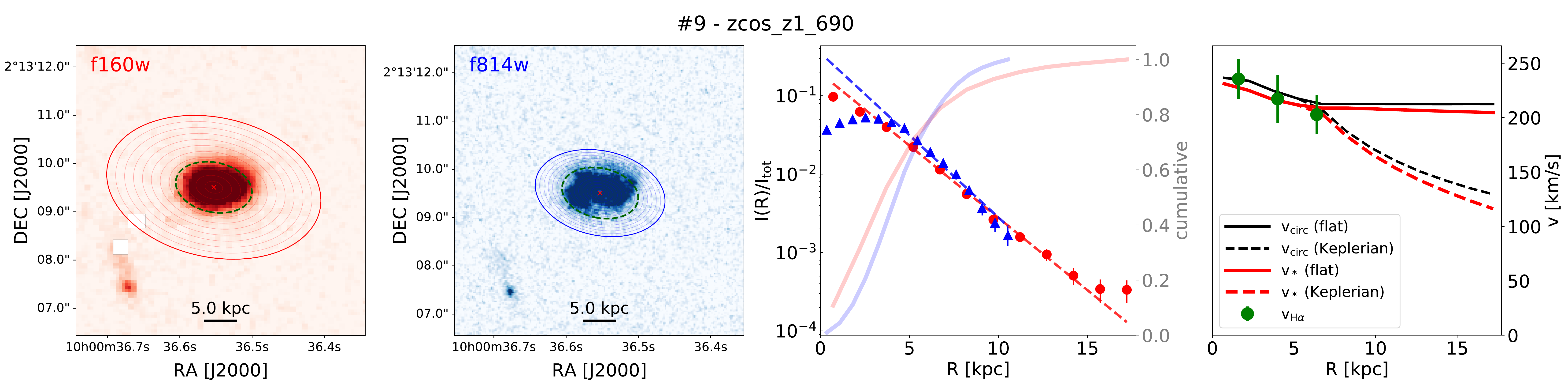}
\includegraphics[width=1.0\textwidth]{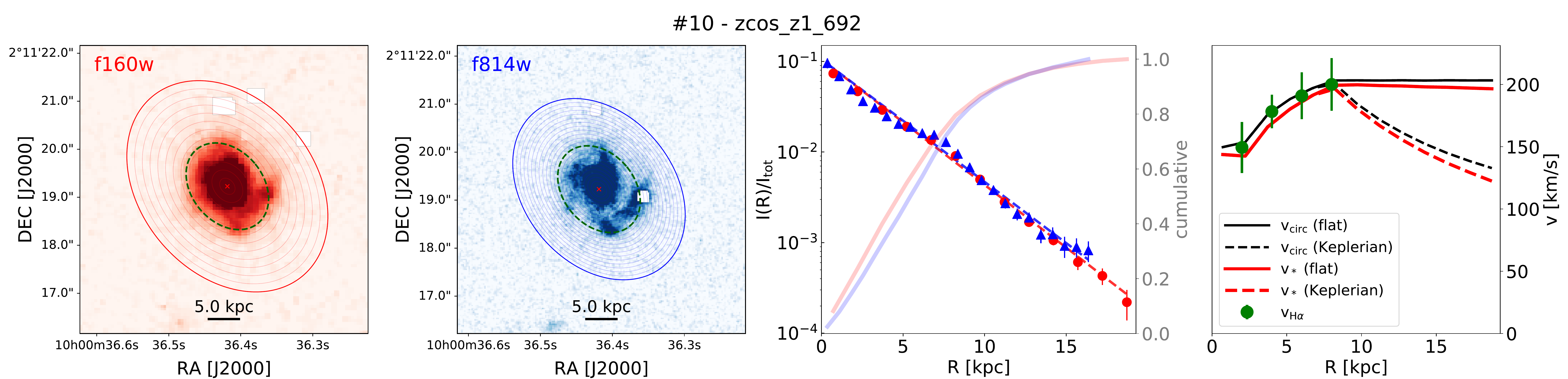}
\includegraphics[width=1.0\textwidth]{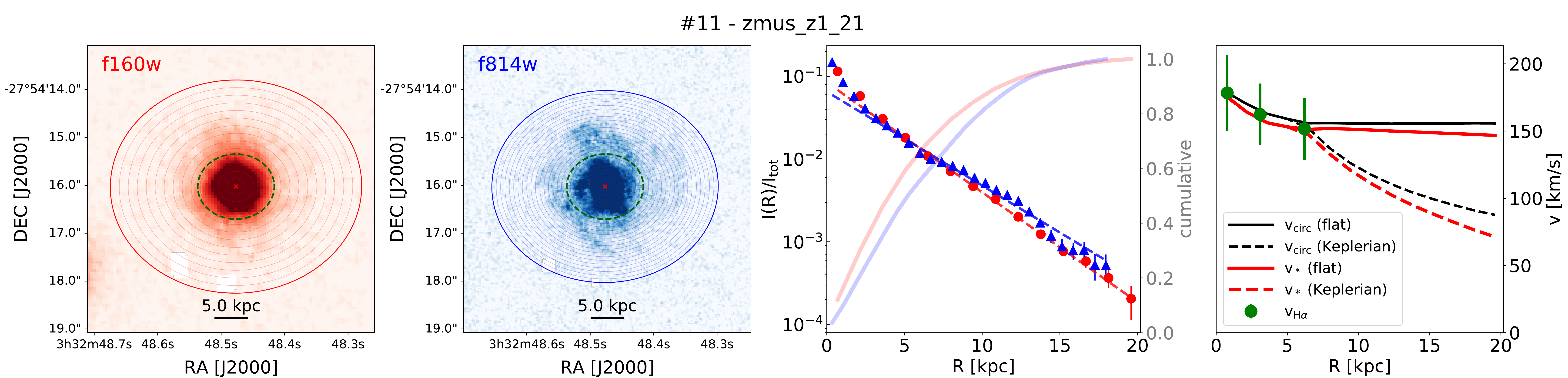}
\caption{Continued}
\end{figure*}
\addtocounter{figure}{-1}

\begin{figure*}[tb]
\centering
\includegraphics[width=1.0\textwidth]{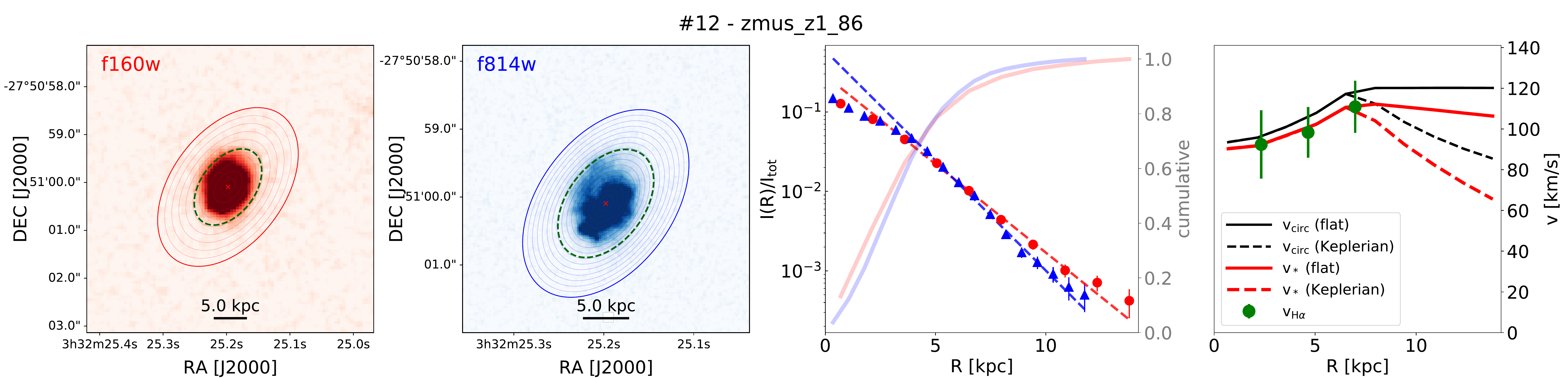}
\includegraphics[width=1.0\textwidth]{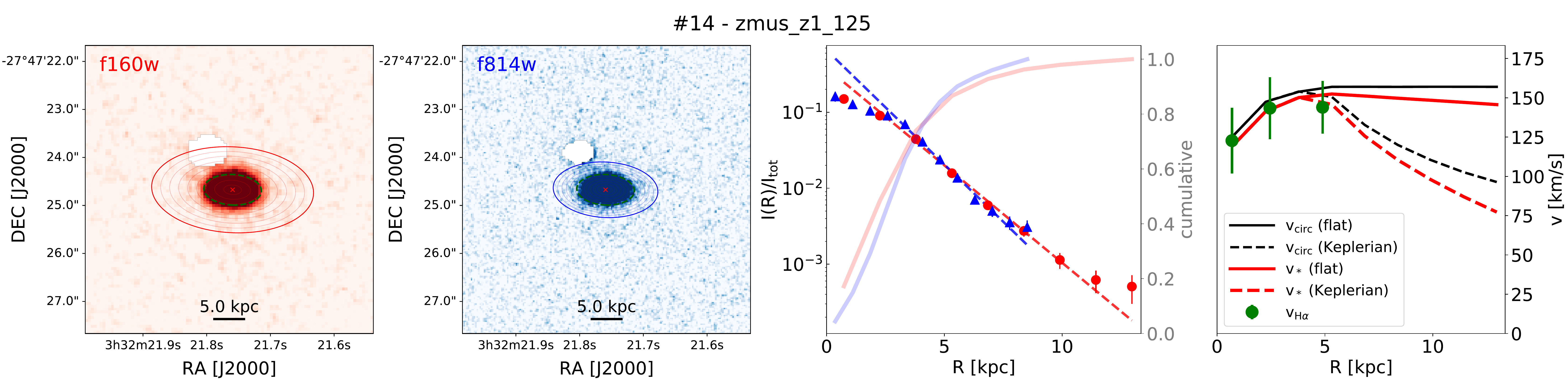}
\includegraphics[width=1.0\textwidth]{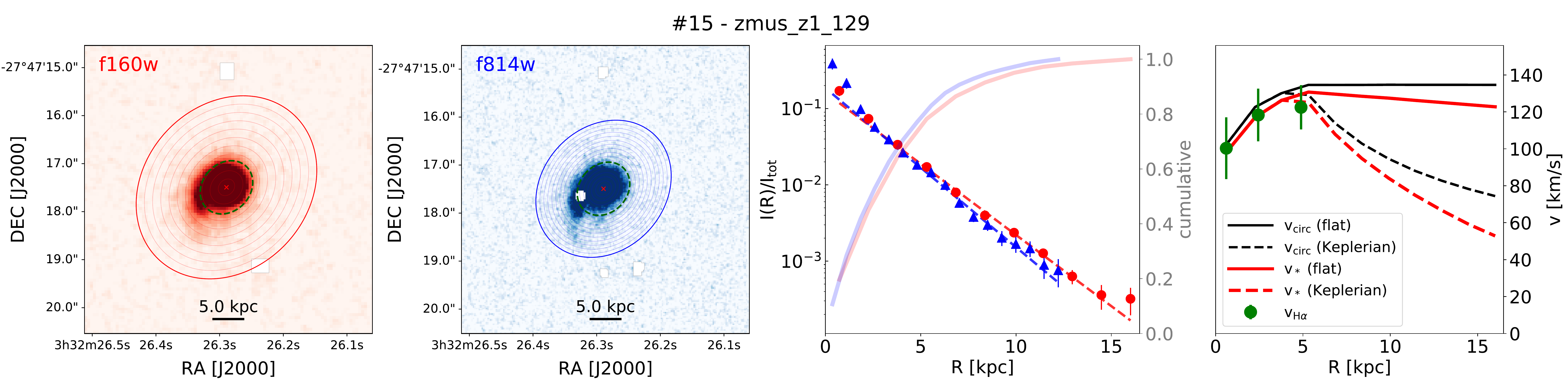}
\includegraphics[width=1.0\textwidth]{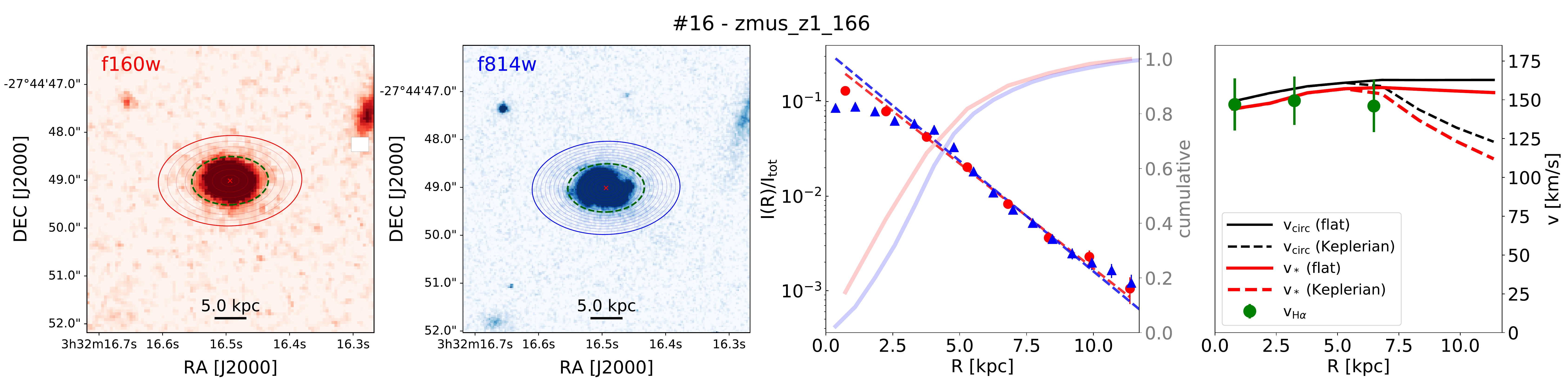}
\includegraphics[width=1.0\textwidth]{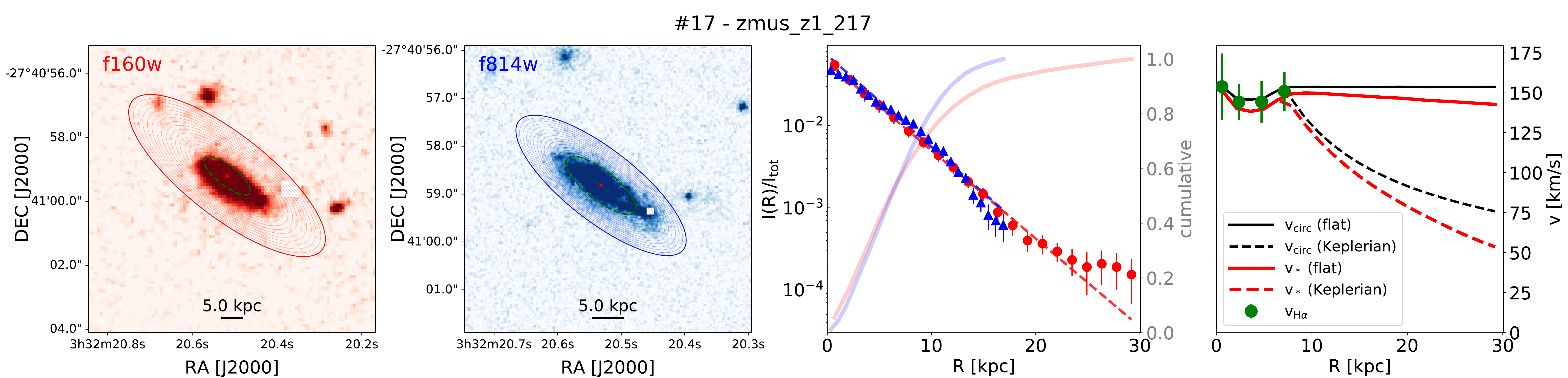}
\caption{Continued}
\end{figure*}
\addtocounter{figure}{-1}

\begin{figure*}[tb]
\centering
\includegraphics[width=1.0\textwidth]{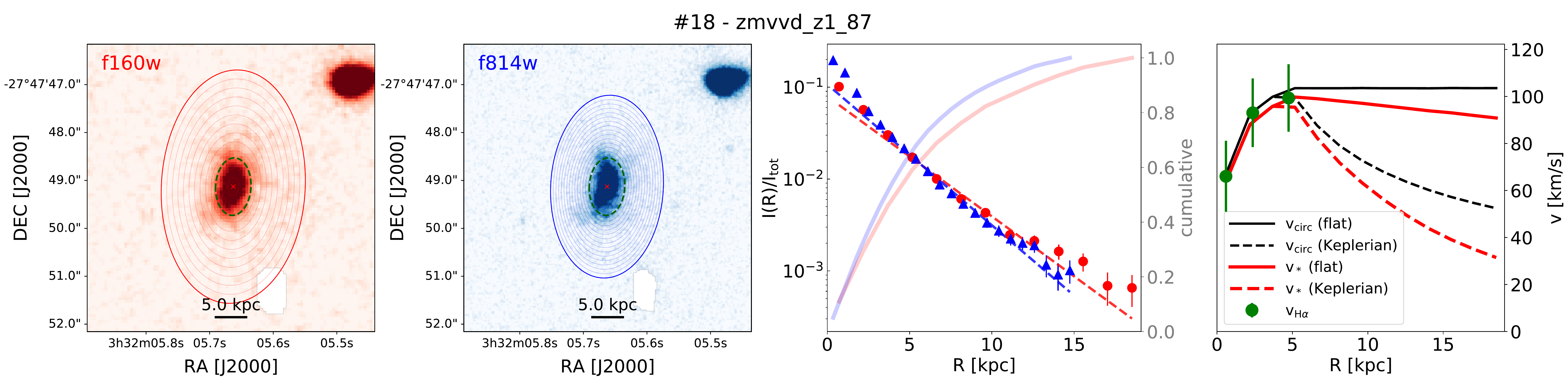}
\caption{Continued}
\end{figure*}

\end{document}